\documentclass[preprint,epsfig,graphicx,showpacs,eqsecnum,aps,amsmath,mathrsfs,eufrak,amssymb,floatfix]{revtex4}
\usepackage{graphicx}
\usepackage{epsfig}
\begin{document}

\title{Semi-microscopic description of the double backbending in some deformed even-even rare earth nuclei}
\author{R. Budaca $^{a)}$ and A. A. Raduta$^{a,b)}$}

\affiliation{$^{a)}$Institute of Physics and Nuclear Engineering, Bucharest,
POB MG6, Romania}
\affiliation{$^{b)}$Academy of Romanian Scientists, 54 Splaiul Independentei, Bucharest 050094, Romania}
\date{\today}
\begin{abstract}
A semi-microscopic model to study the neutron and proton induced backbending phenomena in some deformed even-even nuclei from the rare earth region, is proposed. The space of particle-core states  is defined by the angular momentum projection of a quadrupole deformed product state. The backbending phenomena are described by mixing four rotational bands, defined by a set of angular momentum projected states, and a model Hamiltonian describing a set of paired particles moving in a deformed mean field and interacting with a phenomenological deformed core.  The ground band corresponds to the configuration where all particles are paired while the other rotational bands are built on one neutron or/and one proton broken pair. Four rare earth even-even nuclei which present the second anomaly in the observed moments of inertia are successfully treated within the proposed model.
\end{abstract}
\maketitle

\renewcommand{\theequation}{1.\arabic{equation}}
\section{Introduction}
\label{sec:level1}

The irregular behavior of the moment of inertia in the yrast band at intermediate and high spin states, known as backbending, has always attracted considerable experimental and theoretical attention. Since its first experimental observation \cite{Johnson}, many endeavor attempts were performed in order  to explain the phenomenon. It is commonly accepted that it is caused by the intersection of two rotational bands. This interpretation was proposed by Stephens and Simon \cite{Stephens} based on the rotational alignment of the individual single-particle angular momenta of a broken pair along the rotation axis. The pair breaking is caused by the Coriolis force which violates the time-reversal symmetry. The first theoretical interpretation based on the Coriolis anti-pairing effect was due to Mottelson and Valatin \cite{Mottelson} where the backbending phenomenon was put on the account of a drastic change in the pairing field.

Although the band hybridization method was all along known and applied to this particular problem within some phenomenological approaches \cite{Molinari,Broglia1,Broglia2}, the nature of the involved rotational bands was not yet well established. Only after the rotational alignment hypothesis was confirmed, it became clear that the first backbending is due to the intersection of the ground band ($g$) and a two quasiparticle ($2qp$) band built upon a broken pair from a high angular momentum orbital. The second band is often referred to as the $S$(tockholm) band. Thus, the anomalous increase of the moment of inertia is interpreted as the reduction of the energy cost to achieve a certain total angular momentum by aligning the angular momenta carried by the constituents of a broken pair. Stephens and Simon noticed that in the rare earth region the first broken pair is from the neutron intruder orbital $6i_{13/2}$. Actually this picture  was later confirmed by many theoretical calculations, mostly based on the cranking Hartree-Fock-Bogoliubov (CHFB) \cite{Beng1,Faessler2} calculations and the core plus quasiparticle models \cite{Stephens,Beng2,Grum}. The backbending is a relatively widespread phenomenon within the rare earth region, but only very few nuclei  exhibit a second anomaly in the moment of inertia. It was for the first time measured for $^{158}$Er \cite{Lee}, and the early interpretation was based on the alignment of the individual angular momenta resulting from breaking a $5h_{11/2}$ proton pair \cite{Faessler1}. Other nuclei which exhibit a second moment of inertia anomaly are  located around the $N=90$ rare earth isotopes. The proton nature of the second broken pair was at a first glance queried in Ref.\cite{Bryski}, since in the same energy region of the spectrum, the alignment of a $5h_{9/2}$ neutron broken pair might also play an important role. However, the proton nature of the second backbending was later confirmed by several more detailed theoretical studies \cite{Plosz,Beng1} based on  blocking arguments offered by the experimental investigations of the odd-proton and odd-neutron neighboring nuclei of the $N\approx90$ isotopes \cite{Riley,Holz}. As a result, the second backbending is regarded as being caused by a successive breaking of a neutron and a proton pair, where the neutron broken pair is the one which causes the first backbending \cite{Hara1}. As a matter of fact, the suspected neutron pair $5h_{9/2}$ which may break at a time with $5h_{11/2}$ proton pair  is causing, indeed,  a third anomaly in the moment of inertia of some isotopes of Yb \cite{Li}. Indeed, for this nucleus a weak up-bending  is noticed at spins beyond $J=36$.

Of course band hybridization is a conventional name used in the early publications quoted above.
Hereafter we replace it by band mixing which might be used in a brighter context.

The most extensive calculations on the double backbending were performed in the framework of the CHFB approach, which provided one of the most reliable qualitative description of the phenomenon over a large number of nuclei. One of the most important features of the CHFB approach is that it embraces all the mechanisms known to cause the backbending, that is the particle alignment, the pairing phase transition and the sudden change of deformation. However, the CHFB description is a semiclassical one, which encounters difficulties in describing the states near the band crossing.
An important improvement is obtained by the angular-momentum-projected Tamm-Dancoff approximation which was successfully applied for the dysprosium isotopes \cite{Egido1,Egido2}. Therefore, in order to achieve a quantitative description of the multiple backbending, a full quantal formalism is necessary. Such models were proposed based on mainly two directions: genuine shell model formalisms \cite{Hara1} can trace better the influence of the single-particle degrees of freedom on the pair breaking process while the particle-core models \cite{Ikeda1,Grum} put emphasis on the rotational alignment description. The calculations based on the interacting boson model \cite{Iachello3,Bes} can be also included in the first category. For a quantitative description of the energy spectra with double backbending one advocates for the second solution. The advantage of the particle-core approach consists in the fact that it treats the single-particle and collective degrees of freedom on equal footing.
It is worth mentioning that a qualitative explanation of the first backbending in some isotopes of Pt, W and Os, was obtained in Ref.\cite{Hess} by using the general collective model 
\cite{Gneus1,Gneus2} where, of course, the particle degrees of freedom are missing. Therein the backbending is determined by the angular momentum dependence of the moment of inertia, induced by the specific ways the structure coefficients are fixed.

In a previous publication \cite{Rabura}, we proposed a semi-microscopic model for the description of the backbending phenomenon within the band mixing picture. The rotational bands implied in the  mixing procedure were defined by angular momentum projection from quadrupole deformed product states and a model Hamiltonian describing a set of intruder neutrons interacting among themselves through pairing forces and coupled to a phenomenological deformed core. By projecting the angular momentum  one avoids the difficulties showing up when one treats observables which are sensible to the angular momentum fluctuations. Indeed, working with states of good angular momentum is more advantageous than applying cranking methods which encounter enormous angular momentum fluctuations in the band crossing region. The distinctive feature of our model is that, although we use a spherical projected particle-core basis, the core and the single-particle trajectories are deformed. The mixing of the rotational bands was achieved by diagonalizing the model Hamiltonian in an orthogonal basis constructed from the projected states of $g$ and $S$-bands. The model was meant to reproduce only the first backbending, which was done quite well for six even-even nuclei from the rare earth region. Besides the reproduction of the backbending plots, the formalism  \cite{Rabura} also provided some useful information regarding the rotational alignment of the particles moving in an intruder orbital.

In the present paper we extend the formalism from Ref.\cite{Rabura} to the second backbending induced by a proton broken pair. This is done by performing the mixing of four rotational bands. The first two are obviously the $g$-band and the $S$-band with a neutron broken pair, whereas the other two are associated to a proton broken pair and to two, one of neutron and one of proton type, broken pairs, respectively. The projected states which define the four bands have specific single-particle factors describing each case mentioned above. The protons and neutrons are treated through BCS model states associated only to $6\nu i_{13/2}$ and $5\pi h_{11/2}$ orbitals. The intruder particles are coupled to a phenomenological core which is deformed and described by means of the coherent state model (CSM) \cite{Raduta1}. The projected states are deformed and therefore not orthogonal but can be used to construct an orthogonal basis. The lowest eigenvalues of the model Hamiltonian in this orthogonal basis define the yrast band. The main purpose of the present work is to reproduce the experimental yrast spectrum and its backbending behavior for some even-even rare earth nuclei which are  known to be double backbenders, as well as to provide a through out analysis of the rotational alignment process and the possible consequences for the $E2$ transition properties along the yrast band.

The description of the method and results are presented according to the following plan.
 The model Hamiltonian as well as the projected particle-core product basis used for the description of the double backbending phenomenon is presented in the next section, Section II.The $E2$ transition probabilities are considered in Sec. III and the emerging numerical calculations are given in Sec. IV. Final conclusions are drawn in Sec. V.
\renewcommand{\theequation}{2.\arabic{equation}}
\section{The model Hamiltonian and a particle-core product basis}
\label{sec:level2}

In this study we present a new and simple semi-phenomenological model to explain the first two backbendings seen in some rare earth nuclei. 
The spectra exhibiting a double backbending will be described by a particle-core Hamiltonian whose eigenvalues are calculated within a particle-core space. The nucleons are moving in a deformed mean field and the alike ones interact among themselves by pairing force. The core is deformed and described by a phenomenological quadrupole coherent state \cite{Raduta1}:
\begin{equation}
\psi_{c}=e^{d(b_{20}^{\dagger}-b_{20})}|0\rangle_{b},
\label{coh}
\end{equation}
where $b^{\dagger}_{2\mu}$ with $-2\le \mu \le 2$ denotes the quadrupole boson operator, while $d$ is a real parameter which simulates the nuclear deformation.
The two subsystems interact with each other by a $qQ$  and a spin-spin, $\vec{J}_f\cdot\vec{J}_c$, interaction. The associated Hamiltonian is:  
%%%%%%%%%%%%%%%%%%%%%%%%%%%%%%%%%%%%%%%%%%%%%%%%%%%%%%%%%%%%%%%%%%%%%%%%%%%%%%%%%%%%%%%%%%%%%%
\begin{equation}
H=H_{c}+H_{sp}+H_{pair}+H_{pc}.
\label{H}
\end{equation}
The core term $H_{c}$  is a quadratic polynomial of the quadrupole boson number operator, $\hat{N}=\sum_{\mu}b^{\dagger}_{2\mu}b_{2\mu}$:
\begin{equation}
H_{c}=\omega^{b}_{0}\hat{N}+\omega^{b}_{1}\hat{N}^{2}.
\end{equation}

As for the single-particle Hamiltonian $H_{sp}$, this is a sum of two terms corresponding to neutrons and protons, each of them describing a set of particles in an intruder spherical shell model orbital $|nlj\rangle$:
\begin{equation}
H_{sp}=\sum_{i=\nu ,\pi}(\varepsilon_{n_{i}l_{i}j_{i}}-\lambda_{i})\sum_{m_{i}=all}c_{n_{i}l_{i}j_{i}m_{i}}^{\dagger}c_{n_{i}l_{i}j_{i}m_{i}}.
\end{equation}
Here $c_{nljm}^{\dagger}$ and $c_{nljm}$ are the creation and annihilation operators for a particle in a spherical shell model state $|nljm\rangle$ with the energy $\varepsilon_{nlj}$, while $\lambda$ is the Fermi level energy for the system of paired particles. Alike nucleons interact through a pairing force:
\begin{equation}
H_{pair}=-\sum_{i=\nu, \pi}\frac{G_{i}}{4}P^{\dagger}_{j_{i}}P_{j_{i}},
\end{equation}
where $P_{j}^{\dagger}$ and $P_{j}$ denote the creation and annihilation operators of a Cooper pair in the intruder orbital $j$.

The particle-core interaction consists of two terms, the quadrupole-quadrupole ($qQ$) and the spin-spin interaction:
\begin{eqnarray}
H_{pc}&=&H_{qQ}+H_{J_{f}J_{c}},\nonumber\\
H_{qQ}&=&-A_{c}\sum_{i=\nu,\pi}\sum_{\mu,m_{i},m'_{i}}\langle n_{i}l_{i}j_{i}m_{i}|r^{2}Y_{2\mu}|
n_{i}l_{i}j_{i}m'_{i}\rangle c_{n_{i}l_{i}j_{i}m_{i}}^{\dagger}c_{n_il_ij_{i}m'_{i}}
\left[(-)^{\mu}b_{2-\mu}^{\dagger}+b_{2\mu}\right],\nonumber\\
H_{J_{f}J_{c}}&=&C\vec{J}_{f}\cdot\vec{J}_{c}.
\end{eqnarray}
Here the total angular momentum carried by protons and neutrons is denoted by:
\begin{equation}
\vec{J}_{f}=\vec{J}_{p}+\vec{J}_{n}.
\end{equation}
The interaction strength $A_{C}$ is taken to be the same for neutrons and protons. The parameters $A_{C}$ and $C$ are free in the present work and therefore are to be fixed by a fitting procedure.

The mean field is defined by averaging $\tilde{H}(=H_{sp}+H_{qQ})$   with the coherent state 
(\ref{coh}), which results in obtaining a single-particle Hamiltonian which is similar to the deformed Nilsson Hamiltonian \cite{Nilsson}. In the first order of perturbation, the energies of the deformed mean field are given by:
\begin{eqnarray}
\varepsilon_{nljm}&=&\varepsilon_{nlj}-4dX_{C}(2n+3)C_{\frac{1}{2}\,0\,\frac{1}{2}}^{j\,\,2\,\,j}C_{m\,0\,m}^{j\,\,2\,\,j},\rm{with},\nonumber\\
X_{C}&=&\frac{\hbar}{8M\omega_{0}}\sqrt{\frac{5}{\pi}}A_{C},
\label{sp1}
\end{eqnarray}
where $n$ is the principal quantum number of the intruder orbital, while $M$ and $\omega_{0}$ are the nucleon mass and the harmonic oscillator frequency. $\varepsilon_{nlj}$ denotes the spherical shell model energies corresponding to the parameters given in Ref.\cite{Ring} i.e.,
\begin{equation}
\hbar\omega_0=41A^{-1/3},\;\;C'=-2\hbar\omega_0\kappa,\;\; D=-\hbar\omega_0\kappa\mu.
\end{equation} 
where the parameters ($\kappa,\mu$) have the values $(0.0637, 0.42)$ for neutrons and 
$(0.0637, 0.6)$ for protons. The second term in the right hand side of Eq.(\ref{sp1}) is obtained by averaging the non-spherical part of the mean field with the spherical shell model  state $|nljm\rangle$. The true eigenvalues of the mean-field would be obtained by diagonalization, when the off-diagonal matrix elements of the deformed term are taken into account. 

Of course one could argue that the single-particle energies with linear dependence on the deformation, look unrealistic. One undesired feature is that the state with $j=1/2$ is not affected  by deformation. Actually, we were aware of this drawback and corrected for it 
\cite{Rad65,Rad69}. Briefly, a quadratic term in $d$ could be obtained for example by adding the second order perturbative correction or by adding the monopole-monopole interaction to the particle-core Hamiltonian and then applying the first order perturbation theory to the unperturbed spherical term, as we actually did in the quoted references. Diagonalizing the mean-field Hamiltonian in the "asymptotic" basis one obtains the Nilsson energies and wave functions.  

We opted for the linear dependence on $d$ for energies and  the deformed basis $|nljm\rangle$
because of advantage of having the angular momentum as good quantum number and that happens  despite the fact that the states are deformed. The mentioned problem of $j=1/2$ does not matter at all here since the intruders have high angular momenta. Moreover, for small deformation the 
single-particle energies approximate reasonably well the Nilsson ones. The pragmatic feature which is worth to be mentioned refers to the fact the the Fermi level in our model corresponds to the sub-state $m$ which is equal to the $\Omega$ associated to the Fermi level from the Nilsson scheme.
This feature gave us the certainty  that the essential ingredient for approaching the backbending behavior, is included.

Pairing correlations with such a deformed basis but in a different context has been also used in Ref.\cite{Bes}.
%%%%%%%%%%%%%%%%%%%%%%%%%%%%%%%%%%%%%%%%%%%%%%%%%%%%%%%%%%%%%%%%%%%%%%%%%%%%%%%%%%%%%%%%%%%%%%%%%
Since only the relative energies  to the Fermi level are involved in the BCS equations, the orbital energy $\varepsilon_{nlj}$ is taken to be zero. Moreover, due to the fact that the quantum numbers $n$ and $l$ do not change within a multiplet we simplify the notation and denote the resulting energies by $\varepsilon_{jm}$.  
From here it is obvious that two states related by a time-reversal transformation have the same energy, and therefore one can restrict the single-particle space to the states $|jm\rangle$ with $m>0$, keeping in mind that each such state is occupied by a pair of nucleons.
The sum of the mean field term and the pairing interaction for alike nucleons is brought to a diagonal form through the Bogoliubov-Valatin (BV) transformation:
\begin{eqnarray}
\alpha_{jk}^{\dagger}&=&U_{jk}c_{jk}^{\dagger}-V_{jk}(-)^{j-k}c_{j-k},\nonumber\\
\alpha_{jk}&=&U_{jk}c_{jk}-V_{jk}(-)^{j-k}c_{j-k}^{\dagger}.
\label{BV}
\end{eqnarray}
The output of the BCS calculation consists of the occupation probabilities of the $m$-substates, the gap parameter $\Delta$, as well as the Fermi energy $\lambda$. Consequently the average number of nucleons in the $j$-multiplet, 2$\langle N^{\tau j}_{pair}\rangle$, with $\tau =\nu, \pi$ is readily obtained:
\begin{equation}
\langle N^{\tau j}_{pair}\rangle=\sum_{m>0}V_{\tau jm}^{2}.
\label{pairs}
\end{equation}
 For the chosen nuclei the Fermi levels for neutrons and protons, lie close to a sub-state of the intruders $6i_{13/2}$ and $5h_{11/2}$ respectively. If the particle-core  basis was a deformed one, then the lowest state $|2qp\rangle|\psi_c\rangle$ would correspond to a sub-state of the two intruders, respectively.  The mentioned substates have $m=1/2$ for neutrons and $m=7/2$ for protons. Since the core state does not contribute to the total $K$ quantum number, the projection of the total angular momentum on the symmetry axis, we say that the intrinsic states leading to the yrast band have a $K=1/2$ for the neutrons and $K=7/2$ for protons. Also in the Nilsson model, the last filled neutron state has $\Omega=1/2$ while the last proton occupies the state $\Omega=7/2$.
The choice of the $K=1/2$ sub-state as the Fermi level of the neutron system was made in 
Ref.\cite{Rabura} to describe  the first backbending.  As for  considering the $K=7/2$ Fermi level for the proton system,  breaking the corresponding pair and aligning the resulting quasiparticle angular momenta to that of the core as prerequisite conditions of the second backbending, these features are in full agreement with the microscopic formalism used in literature. In this respect
in Ref.\cite{Beng1} the alignment of a $\Omega=7/2$ broken pair is used to explain the second backbending in $^{158}$Er and $^{160}$Yb. The last nucleus mentioned is also treated by Cwiok and collaborators  in Ref.\cite{Cwiok} while $^{158}$Er by Riley \cite{Riley}.   

A great simplification is obtained if the single-particle space is restricted to the intruder multiplets  where a number of nucleons equal to 2$\langle N^{\tau j}_{pair}\rangle$ is distributed.
Solving the BCS equations in the restricted space, the quasiparticle energies depend on $m$ but are still invariant at changing $m$ to $-m$. However, in  a pure microscopic formalism where the Coriolis interaction is included in the  mean field, the time reversed quasiparticle states are no longer degenerate and consequently the broken pair is a $K=1$ state. Here the term $\vec{J_f}\cdot\vec{J}_c$, which simulates the Coriolis interaction in the sense specified in Ref.
\cite{Rabura} is only subsequently used, when the whole Hamiltonian is diagonalized and thereby the broken pairs with $K=1$ are used. An important technical simplification is achieved if these pairs are obtained by applying the angular momentum raising operator on the $K=0$ pairs. 

If the quasiparticles were not deformed and moreover the dangerous graphs were eliminated at the level of BCS calculations, one  would expect that the interaction between states with different number of quasiparticles  is vanishing. Under these circumstances, truncating the particle-core space to the states with $0qp, 2qp$ and $4 qp$ is a reasonable approximation. Since the rotation process involved in the angular momentum projection operation changes the $K$ quantum number, and moreover particles and holes are mixed by the BV transformation,  the overlap of states with different number of particles is however nonvanishing. Despite this feature we keep the restriction of the quasiparticle space as specified above. The reason is that the mixing weight of components with more than 4 quasiparticles would be at least of sixth order in the $U$ and $V$ coefficients and consequently small.

Thus, the restricted space of angular momentum projected states to be used for treating the model Hamiltonian, $H_{qp}$, written in the quasiparticle representation is:

\begin{equation}
\left\{\Psi_{JM}^{(1)},\Psi_{JM;1}^{(2)}(j_{n}\nu),\Psi_{JM;1}^{(3)}(j_{p}\pi),\Psi_{JM;2}^{(4)}(j_{n}\nu;j_{p}\pi)\right\}.
\label{basisset}
\end{equation}
The set members are defined by:
\begin{eqnarray}
\Psi_{JM}^{(1)}&=&\mathcal{N}_{J}^{(1)}P_{M0}^{J}|nBCS\rangle_{d}|pBCS\rangle_{d}\psi_{c},
\label{st1}\\
\Psi_{JM;1}^{(2)}(j_{n}\nu)&=&\mathcal{N}_{J1}^{(2)}(j_{n}\nu)P_{M1}^{J}
\left[J_{+}\alpha_{j_{n}\nu}^{\dagger}\alpha_{j_{n}-\nu}^{\dagger}|nBCS\rangle_{d}\right]|pBCS\rangle_{d}\psi_{c},\label{st2}\\
\Psi_{JM;1}^{(3)}(j_{p}\pi)&=&\mathcal{N}_{J1}^{(3)}(j_{p}\pi)P_{M1}^{J}
|nBCS\rangle_{d}\left[J_{+}\alpha_{j_{p}\pi}^{\dagger}\alpha_{j_{p}-\pi}^{\dagger}|pBCS\rangle_{d}\right]\psi_{c},\label{st3}\\
\Psi_{JM;2}^{(4)}(j_{n}\nu;j_{p}\pi)&=&\mathcal{N}_{J2}^{(4)}(j_{n}\nu;j_{p}\pi)P_{M2}^{J}
\left[J_{+}\alpha_{j_{n}\nu}^{\dagger}\alpha_{j_{n}-\nu}^{\dagger}|nBCS\rangle_{d}\right]
\left[J_{+}\alpha_{j_{p}\pi}^{\dagger}\alpha_{j_{p}-\pi}^{\dagger}|pBCS\rangle_{d}\right]\psi_{c},\nonumber\\
\label{st4}
\end{eqnarray}
where the reciprocal norms can be analytically expressed. Also, $\alpha^{\dagger}_{j\mu}/\alpha_{j\mu}$ stand for the creation/annihilation  quasiparticle operators. 
The Hill-Wheeler projection operator \cite{Hill} has the form:
\begin{equation}
P_{MK}^{J}=\frac{2J+1}{8\pi^{2}}\int D_{MK}^{J*}\hat{R}(\Omega)d\Omega.
\end{equation}
The angular momentum projection from the many body fermion states is achieved by using the procedure of Ref.\cite{Kelemen}.
 The Pauli principle restrains the maximal angular momentum of a given configuration \cite{Hamermesh} to
\begin{equation}
J^{max}_{\tau}=N^{\tau j}_{pair}(2j_{\tau}-2N^{\tau j}_{pair}+1),
\label{Hamer}
\end{equation}
where $N^{\tau j}_{pair}$ pairs of $\tau$ particles, occupy the states of angular momentum 
$j_{\tau}$.

The set of projected states mentioned above, is not orthogonal.
We orthogonalized first the angular momentum projected basis and then diagonalized the model Hamiltonian written in the quasiparticle representation. Note that the bands mixing is achieved by two processes, the orthogonalization procedure of the initial basis and then by diagonalizing the model Hamiltonian $H$ (\ref{H}).
The lowest eigenvalues of the total Hamiltonian $H$ in the orthogonal basis defines the yrast band.

The energy spectrum of the rotational bands  is approximated by the average of the total Hamiltonian with each projected state from the set (\ref{basisset}). 

The mixing of these bands is achieved following the procedure of Ref.\cite{Rabura} extended to the case of four interacting bands. Here we briefly present the main ingredients of this procedure. 

Indeed, denoting by $\alpha_{m}^{J}$ the eigenvalues and by $V_{im}^{J}$  the eigenvectors of the overlap matrix  corresponding to $J\ne 0$, it can be checked that the set of functions
\begin{equation}
\Phi_{m}^{JM}=\frac{1}{\sqrt{\alpha_{m}^{J}}}\sum_{i=1}^{4}\Psi_{JM}^{(i)}V_{im}^{J},\,\,\,\,m=1,2,3,4,
\label{basis}
\end{equation}
is orthogonal. 

Writing the total wave function as an expansion in the newly obtained orthogonal basis:
\begin{equation}
\Phi_{Tot}^{JM}=\sum_{m=1}^{4}X_{m}^{J}\Phi_{m}^{JM},
\label{Totstate}
\end{equation}
 the eigenvalue equation associated to the model Hamiltonian acquires the following matrix form:
\begin{equation}
\sum_{m'=1}^{4}\tilde{H}^{(J)}_{mm'}X_{m'}^{J}=E_{J}^{m}X_{m}^{J}.
\label{eigen}
\end{equation}
The Hamiltonian matrix $\tilde{H}^{(J)}_{mm'}$ is defined as
\begin{equation}
\tilde{H}^{(J)}_{mm'}=\frac{1}{\sqrt{\alpha_{m}^{J}\alpha_{m'}^{J}}}\sum_{n,n'=1}^{4}V_{nm}^{J}\langle\Psi_{JM}^{(n)}|H|\Psi_{JM}^{(n')}\rangle V_{n'm'}^{J}.
\label{Htilde}
\end{equation}

Solving the homogeneous system of linear equations (\ref{eigen}) for a given $J\ne 0$ and then changing $J$, one obtains a four $J$-sets of energies. Collecting the lowest energy from each $J$-set of solutions, one obtains the so called yrast band.

%%%%%%%%%%%%%%%%%%%%%%%%%%%%%%%%%%%%%%%%%%%%%%%%%%%%%%%%%%%%%%%%%%%%%%%%%%%%%%%%%%%%%%%%%%%%%%%%
\renewcommand{\theequation}{3.\arabic{equation}}
\section{$E2$ transition probabilities}
\label{sec: level3}

The reduced quadrupole transition probabilities are calculated by truncating the transition operator
to the boson part, i.e. we suppose that the collective transition is due to the core component of the wave function. The microscopic structure of the yrast states have however an indirect contribution. The boson structure of the transition operator is assumed to be of the form: 
\begin{equation}
Q_{2\mu}=q^{\prime}_{1}\alpha_{2\mu}+q^{\prime}_{2}\left(\alpha\alpha\right)_{2\mu},
\end{equation}
where $\alpha_{2\mu}$ denotes the quadrupole coordinate which is depending linearly on the boson operators
\begin{equation}
\alpha_{2\mu}=\frac{1}{\sqrt{2}}(b_{2\mu}^{\dagger}+(-)^{\mu}b_{2-\mu}).
\end{equation}
In terms of quadrupole bosons the transition operator has the expression:
\begin{equation}
Q_{2\mu}=q_{1}\left(b^{\dagger}_{2\mu}+(-)^{\mu}b_{2, -\mu}\right)+q_{2}\left((b^{\dagger}_2b^{\dagger}_2)_{2\mu}+2(b^{\dagger}_2b_{\tilde {2}})_{2\mu}+(b_{\tilde{2}}b_{\tilde{2}})_{2\mu}\right),
\end{equation}
where:
\begin{equation}
q_{1}=\frac{1}{\sqrt{2}}q'_1, \; q_{2}=\frac{1}{2}q'_2.
\end{equation}
The reduced probability for the quadrupole transition in the yrast band, using the Rose's convention \cite{Rose}, can be written as
\begin{equation}
B(E2,J^{+}\,\rightarrow\,J'^{+})=\left|\langle\Phi^{J}_{Tot}||Q_{2}||\Phi^{J'}_{Tot}\rangle\right|^{2},
\end{equation}
where the functions involved are the  states (\ref{Totstate}) obtained by diagonalizing the matrix
$\tilde{H}^{(J)}_{mm'}$. If the final state is $0^+$, then instead of $\Phi^{J'}_{Tot}$ with $J'=0$ we use $\Psi_{0}^{(1)}$.
The transition matrix elements involve two parameters $q_{1}$ and $q_{2}$,
which are to be fixed by a fitting procedure. The reduced matrix elements of the transition operator have been analytically expressed in Refs. \cite{RadSab,Rad78}.

%%%%%%%%%%%%%%%%%%%%%%%%%%%%%%%%%%%%%%%%%%%%%%%%%%%%%%%%%%%%%%%%%%%%%%%%%%%%%%%%%%%%%%%%%%%%%%%%

\renewcommand{\theequation}{4.\arabic{equation}}
\section{Numerical application and discussions}
\label{sec: level4}

There are very few nuclei in the rare earth region which present a second anomaly in their moment of inertia evolution along the yrast band. The most studied nuclei  are $^{156}$Er, $^{158}$Er, $^{160}$Yb and $^{162}$Hf since for them a great deal of experimental data are available.  These  nuclei will be treated within the formalism described in the previous sections.

\subsection{Parameters}

The model involves seven parameters. Six of them, namely the neutron and proton pairing constants $G_{n}$ and $G_{p}$, the strengths of the $qQ$ and spin-spin interactions, $X_{C}$ and $C$, and the strengths $\omega_{0}^{b}$ and $\omega_{1}^{b}$, of the two  boson terms, are the structure coefficients  defining  the model Hamiltonian. The remaining parameter $d$  defines the coherent state $\psi_{c}$ and plays the role of the deformation parameter. The fitted values of these parameters are given in Table I. In what follows we shall explain how these parameters were fixed.

\begin{table}[h!]
\caption{The fitted parameters for the four nuclei are listed. The nuclear quadrupole deformation $\beta_{2}$, taken from Ref.\cite{Lala}, is presented for comparison with the deformation parameter $d$.}
\vspace{0.2cm}
{\footnotesize
\begin{tabular}{|c|c|c|c|c|c|c|c|c|c|c|c|}
\hline
Nucleus&$d$&$X_{C}$ [keV]&$G_{n}$ [MeV]&$G_{p}$ [MeV]&$\omega_{0}^{b}$ [MeV]&$\omega_{1}^{b}$ [keV]&$C$ [keV]&$d\cdot X_{C}$ [keV]&$\beta_{2}$\\
\hline
$^{156}$Er&1.9498&84.0455&0.2146&0.2626&1.1420&~0.255&~3.042&163.87&0.177\\
$^{158}$Er&2.4910&68.6731&0.1803&0.2593&1.1525&-1.426&~5.866&171.06&0.203\\
$^{160}$Yb&2.2870&74.9940&0.1892&0.2619&1.2684&-0.514&~2.270&171.51&0.195\\
$^{162}$Hf&2.1490&78.2942&0.2000&0.2583&1.3104&~8.674&-1.991&168.25&0.184\\
\hline
\end{tabular}}
\end{table}

In the  first step, the BCS equations were separately  solved for protons and neutrons. The pairing constants and the single-particle energies represent the input data for the BCS equations. The single-particle energies are defined by Eq.(\ref{sp1}) and depend linearly on the deformation parameter, as can be seen from Fig.{\ref{sptot1} and Fig.\ref{sptot2}. From these one can see that the product $dX_{C}$ plays the role of the deformed mean field strength, like the quadrupole nuclear deformation $\beta_{2}$ in the Nilsson model \cite{Nilsson}. Given the fact that here we deal only with neutrons from the $6i_{13/2}$ intruder orbital and protons from $5h_{11/2}$ intruder orbital, which are responsible for the first and  the second band crossing respectively, the BCS equations are solved only for a subset of the entire neutron and proton single-particle space which contains  the states that might interact with the mentioned intruder states. Since the single-particle energies yielded by the deformed mean-field are $m$-dependent quantities, the substates of the intruders will be specified by adding a lower index $m$ to the standard notation specific to the spherical single-particle states. Thus, for neutrons, the subset comprises all states of the $n=5$ shell, excepting the substates with $|m|<11/2$ coming from $5h_{11/2}$ orbital, together with the intruder states $6i_{13/2,m}$ and the state $5h_{11/2,11/2}$ coming from below, which is an intruder for the $n=4$ shell. Similarly, the proton subset comprises all states of the $n=4$ shell, the intruder state 
$4g_{9/2,9/2}$ for the $n=3$ shell coming from below and of course all intruder states $5h_{11/2,m}$. In total, one has to solve the BCS equations in a space of 23 neutron states and 17 proton states where each single-particle state can accommodate two nucleons. The nuclei  $^{158}$Er, $^{160}$Yb and $^{162}$Hf are $N=90$ isotones, such that we distributed in the neutron subspace 10 particles for each, and 20, 22 and 24 particles in the proton subspace respectively. As for 
$^{156}$Er, this has 8 neutrons and 20 protons distributed in the corresponding subspaces. Judging from the observed degree of the shell filling, for all considered nuclei the last occupied proton intruder state $h_{11/2}$ has the projection $7/2$, while the neutron intruder state $i_{13/2}$ which is closest to the neutron Fermi level has the projection $1/2$. Thus, the $m$ substates which correspond to the broken  neutron and proton pairs  $(m_{\nu},m_{\pi})=(1/2,7/2)$ is the same for all four nuclei, even though they have different neutron and proton numbers.

The pairing interaction constants $G_{n}$ and $G_{p}$ and the $qQ$ interaction strength are fixed so that  the observed sequence of the single-particle levels and the last occupied state for a given deformation $d$ of the core are reproduced. Later on, a fine tuning is performed in order to improve the position of the band crossing points. Solving the BCS equations one obtains the quasiparticle energies, the gap parameter $\Delta$, the Fermi level energy $\lambda$ and the occupation probability parameters $U$ and $V$. The projected neutron and proton single-particle states (\ref{st1})-(\ref{st4}) describe only the nucleons from the intruder orbitals $6\nu i_{13/2}$ and $5\pi h_{11/2}$. Thus, in further calculation one would need only the BCS parameters concerning the seven neutron states $i_{13/2}$ and the six proton states $h_{11/2}$. Using the occupation probabilities of the intruder states, one calculates the average number of pairs in the considered intruder orbitals, $\langle N^{\tau j}_{pair}\rangle$.

It is needless to say that the BCS calculations performed only for the single-particle states of the considered intruder orbitals with a number $\langle N^{\tau j}_{pair}\rangle$ of occupying pairs  would provide results equivalent to those obtained for the larger single-particle subspaces chosen above. Even though the equation (\ref{Hamer}) is designed for an even and integer number of pairs, it can be used to determine an approximate higher limit of the angular momentum realized in a virtual configuration of $\langle N^{\tau j}_{pair}\rangle$ pairs. The value obtained in this manner is then rounded to the closest even integer, defining in this way the upper limits of the summations over neutron and proton angular momenta $J_{n}$ and $J_{p}$ involved in the definition of the projected single-particle states. All this information and the BCS results are given in Table II. With all these data, the projected states (\ref{st1})-(\ref{st4}) are fully determined.

Concluding, the BCS calculation in the extended single-particle space is used to calculate the average number of pairs in the intruder orbitals. Once these are determined we solved the BCS equation for each intruder keeping the obtained restriction for the number of the 
$\tau$-particles. Also using the average number of the $\tau$-pairs we calculate the maximal values of 
the angular momentum carried by the given system of fermions. In this way the space of the four particle-core projected states is readily defined. Thus,  we stress again the fact that our method is based on a single $j$ calculation and not on a many $j$. We used the many $j$ calculation just to remove the ambiguity in determining the number of the $\tau$-nucleons which should be distributed among the intruder substates. 

\begin{table}[h!]
\caption{The neutron and proton Fermi level energies, gap parameters and the quasiparticle energies are given for each treated nucleus. The average number of pairs determined with (\ref{pairs}) and the corresponding exact and approximated maximal angular momenta (\ref{Hamer}), obtained by replacing  the number of pairs $N^{\tau j}_{pair}$ by the average number $\langle N^{\tau j}_{pair}\rangle $, are also given.}
{\scriptsize
\begin{tabular}{|c|c|c|c|c|c|c|c|c|c|c|c|c|}
\hline
&\multicolumn{6}{c|}{Neutron $k=1/2$}&\multicolumn{6}{c|}{Proton $k=7/2$}\\ \hline
Nucleus&$\lambda_{n}$ [MeV]&$\Delta_{n}$ [MeV]&$E_{qp}^{n}$ [MeV]&$\left\langle N_{pair}^{\nu i_{13/2}}\right\rangle$&$\langle J_{n}^{max}\rangle$&$J_{n}^{max}$&$\lambda_{p}$ [MeV]&$\Delta_{p}$ [MeV]&$E_{qp}^{p}$ [MeV]&$\left\langle N_{pair}^{\pi h_{11/2}}\right\rangle$&$\langle J_{p}^{max}\rangle$&$J_{p}^{max}$\\
\hline
$^{156}$Er&48.350&1.39475&1.44662&0.95&11.51&12&44.271&1.46021&1.46081&3.33&17.79&18\\
$^{158}$Er&48.496&1.13589&1.13905&1.26&14.50&14&44.083&1.39234&1.39288&3.34&17.77&18\\
$^{160}$Yb&48.268&1.22848&1.23084&1.29&14.74&14&44.395&1.35984&1.46115&3.71&16.98&16\\
$^{162}$Hf&48.056&1.34656&1.34681&1.30&14.78&14&44.678&1.27407&1.61959&4.05&15.81&16\\
\hline
\end{tabular}}
\end{table}

\begin{figure}[ht!]
\begin{center}
\includegraphics[width=1\textwidth]{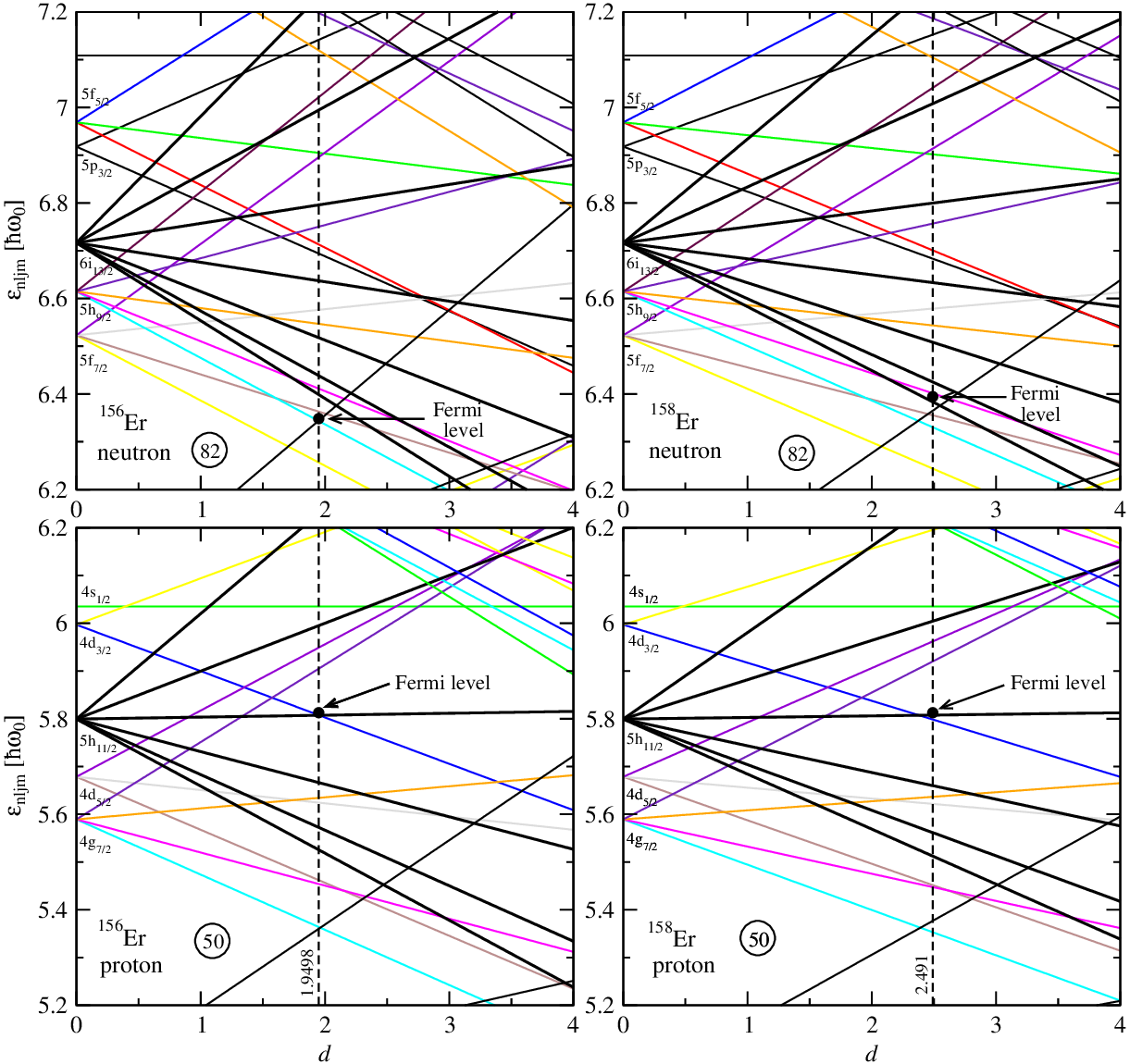}
\end{center}
\caption{Neutron (top) and proton (bottom) single-particle energy levels given in units of $\hbar\omega_{0}$($=41A^{-1/3}$ MeV) and calculated with Eq.(\ref{sp1}) for $^{156}$Er (left) with $X_{C}=84.0455$ keV and $^{158}$Er (right) with $X_{C}=68.6731$ keV. The vertical lines indicate the single-particle configurations corresponding to the fitted deformation parameter $d$. The Fermi energy level resulting form the BCS calculations is also pointed out.}
\label{sptot1}
\end{figure}

\begin{figure}[ht!]
\begin{center}
\includegraphics[width=1\textwidth]{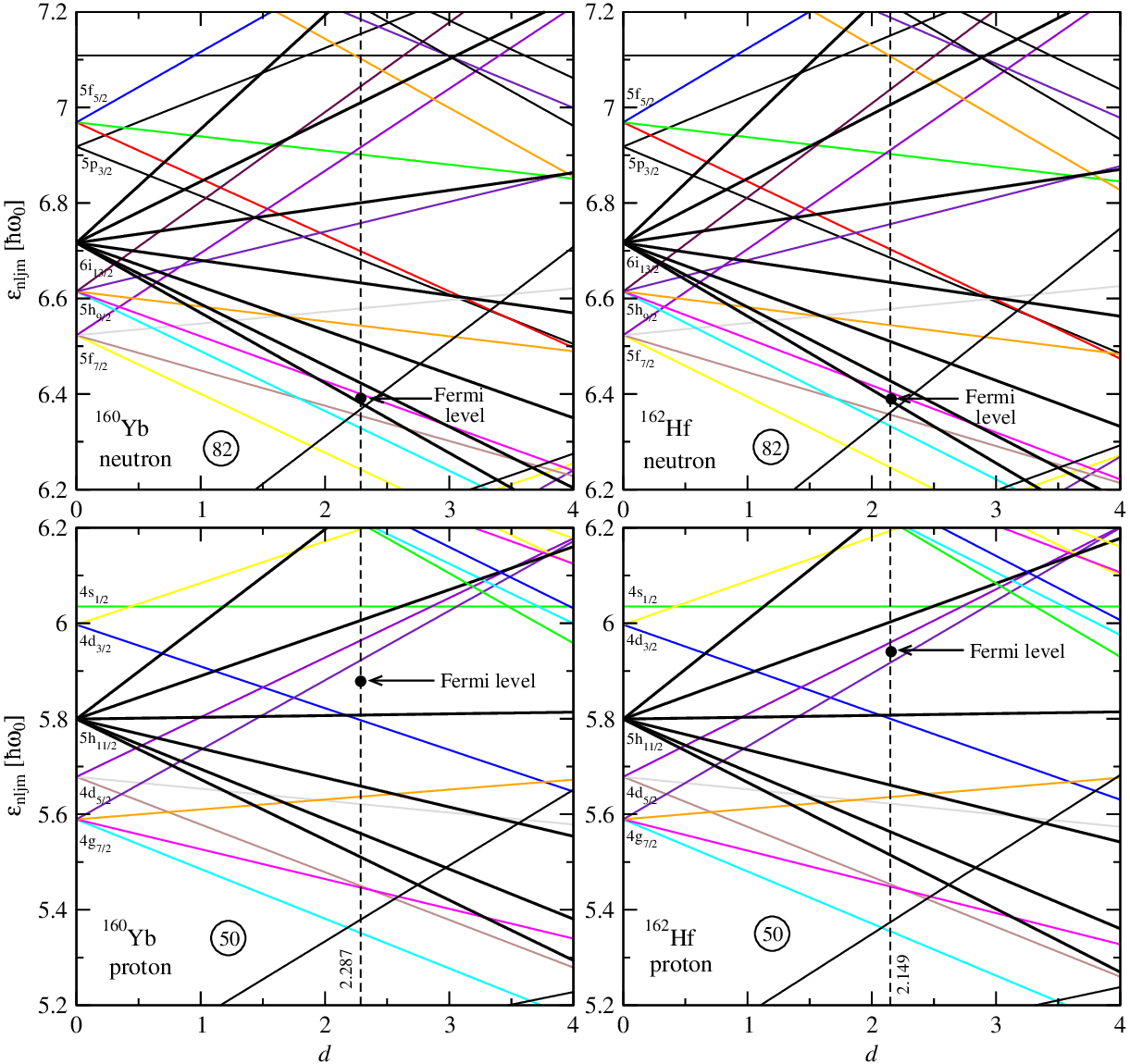}
\end{center}
\caption{Neutron (top) and proton (bottom) single-particle energy levels given in units of $\hbar\omega_{0}$($=41A^{-1/3}$ MeV) and calculated with Eq.(\ref{sp1}) for $^{160}$Yb (left) with $X_{C}=74.9940$ keV and $^{162}$Hf (right) with $X_{C}=78.2942$ keV. The vertical lines indicate the single-particle configurations corresponding to the fitted deformation parameter $d$. The Fermi energy level resulting form the BCS calculations is also pointed out.}
\label{sptot2}
\end{figure}
%%%%%%%%%%%%%%%%%%%%%%%%%%%%%%%%%%%%%%%%%%%%%%%%%%%%%%%%%%%%%%%%%%%%%%%%%%%%%%%%%%%%%%%%%%%%%%%%%%
Some remarks concerning the BCS results are worth to be made. The observed single-neutron level structure of all four nuclei for the tabulated values of the quadrupole deformation $\beta_{2}$ shows that none of the neutron intruder states $i_{13/2}$ are occupied. In the present model, the single-particle energies (\ref{sp1}) depend linearly on the deformation parameter $d$ contrary to the Nilsson case. Because of this feature, one finds that for the $N=90$ isotones the Fermi level provided by the BCS equations is right above the first intruder state $6\nu i_{13/2}$, as  indicated in figures \ref{sptot1} and \ref{sptot2}. Exception is for the $^{156}$Er isotope which has fewer neutrons and  cannot fill any intruder $6\nu i_{13/2}$ sub-state, but due to the large value of the pairing strength $G_{n}$ the occupation probability is considerably extended above the Fermi level $\lambda_{n}$ and thus placing an average number of two nucleons in the intruder orbital $i_{13/2}$ (see Table II). Also, according to the shell filling, the proton Fermi level of the Er isotopes must be placed under the intruder state $5\pi h_{11/2,7/2}$, but as can be seen in Fig.\ref{sptot1} the obtained Fermi level $\lambda_{p}$ is placed right above this state. This is caused by the fact that the intruder state $h_{11/2,7/2}$ and the state $d_{3/2,1/2}$ of the $n=4$ shell are almost degenerated for the chosen value of the deformation parameter $d$ and consequently the occupation probability corresponding to one pair of protons is shared by the two states.

\begin{figure}[th!]
\begin{center}
\includegraphics[width=0.6\textwidth]{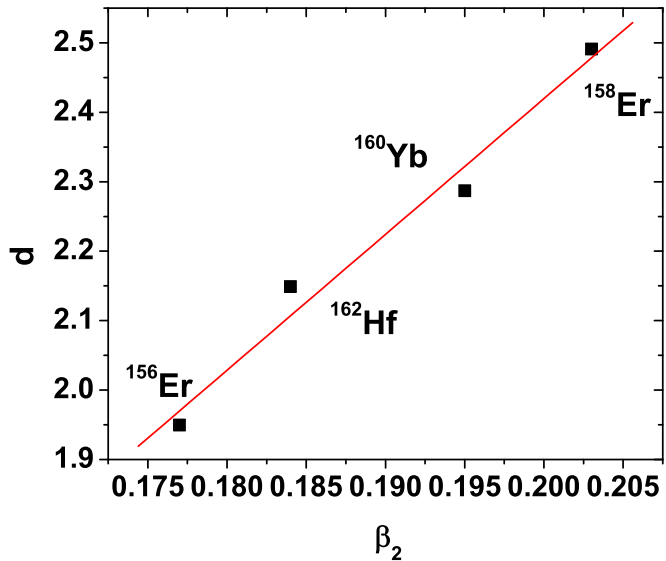}
\end{center}
\caption{ The deformation parameter $d$ (square) is presented as function of the nuclear deformation $\beta_2$. The fitted values of $d$ can be interpolated by the straight line given by 
$d=19.55\beta_2-1.49$.}
\label{dvsb}
\end{figure}
The deformation parameter $d$ affects both the single-particle and the collective degrees of freedom. Indeed,  on one hand it is embedded in the strength $dX_{C}$ of the deformed mean field, and on the other hand it defines the energy of the core. One may therefore assert that the particle-core interaction induces a deformation effect on both the single-particle and the core motion. However, it can be easily checked that the ground band energies are not very sensible to the single-particle degrees of freedom. Indeed, the overwhelming contribution to the total energy of the ground band is due to the core because all the intruder particles are paired and do not carry any angular momentum. This fact implies that up to the first band crossing the whole angular momentum dependency is given by the core. 

Besides the deformation parameter $d$, the core energy is also parametrized by $\omega_{0}^{b}$ and $\omega_{1}^{b}$, the strengths of the two boson terms. The core parameters $d$, $\omega_{0}^{b}$ and $\omega_{1}^{b}$ are determined in the first approximation such that to reproduce the first yrast energy levels which are purely collective. The final value of the deformation $d$ is fixed by achieving a consensus between the reproduction of the single-particle levels configuration and the best description of the angular momentum dependency of the total energy of the $g$-band up to the first backbending.

The final touch to the formalism is made by fixing the strength $C$ of the spin-spin interaction. The effect of the spin-spin interaction was presented in detail in Ref.\cite{Rabura}$^1$
\setcounter{footnote}{1}\footnotetext{In Table II of Ref.\cite{Rabura} the values of $C\cdot 10$ were listed. By a lamentable error the factor 10 accompanying $C$ was omitted. However, all results  of the quoted reference correspond to the true values of $C$.}. Basically, it simulates the Coriolis force in the intrinsic reference frame and is actually the model Hamiltonian term which is responsible for the pair breaking. Indeed, recalling the fact that the   pair breaking is equivalent to the time-reversal symmetry breaking of the system it is then clear that it cannot be achieved by the $qQ$ interaction and therefore the spin-spin interaction is necessarily demanded. It is found that this term  does not have any effect on the energies of  crossing bands up to the first critical angular momentum, but on the contrary  has a strong effect on the moderate and high spin states in the yrast band. Because of this feature the strength $C$ is fixed such that to reproduce the moderate and high spin yrast state energies.

Apparently the number of parameters used is large, but three of them, $d$, $G_p$ and $G_n$, are not freely changed when we pass from one nucleus to another. Indeed, the deformation parameter depends linearly on the nuclear deformation and therefore fixing it for one nucleus, for example by fitting the $B(E2;0^+\to 2^+)$ value, it is known for all remaining ones. This is shown in Fig. 3, where the fitted values of $d$ were interpolated by a straight line.

Also, the results for the strengths of  proton and neutron pairing interactions can be interpolated by a function  proportional with $1/A:$
\begin{equation}
G_p=\frac{41.410}{A}\;{\rm MeV},\;\;G_n=\frac{31.165}{A}\;{\rm MeV}.
\end{equation}
The $A$ dependence of $G_{\tau},\; \tau=p,n$ is quite close to the $A$-parametrization of the interaction strengths which interpolates the values obtained by fitting the even-odd mass difference:
\begin{equation}
G_p=\frac{42.316}{A}\;{\rm MeV},\;\;G_n=\frac{31.360}{A}\;{\rm MeV}.
\end{equation}

\subsection{Energies}

The energies of the rotational bands implied in the present model are approximated by the diagonal matrix elements of the model Hamiltonian between the projected states of the set (\ref{basisset}) and calculated using the parameters listed in Table I. The band mixing  is achieved by orthogonalizing
the projected states (\ref{basisset}) and then diagonalizing $H$ (\ref{H}) in the resulting orthogonal basis. For a given total angular momentum $J$  one obtains a set of four eigenvalues $E_{J}^{m}$, with $m=1,2,3,4$. The lowest energies $E_{J}^{m}$ define the yrast band $E(J)$.

The band mixing is schematically shown in Fig.\ref{etot} where all involved rotational bands and the resulting yrast band are plotted versus total angular momentum $J$. Similar dependence of the rotational bands on the angular momentum was obtained in Ref.\cite{Stephens,Hara1} where the energies were computed only in a projected quasiparticle space with a relatively large number of single-particle states. As can be seen from Fig.\ref{etot} the proton $S$-band does not interact with the other bands or is very weakly interacting with the $g$-band at high spin states in the case of $^{162}$Hf. Moreover, its energy is higher than that of other bands, such that it has no influence on the yrast band. Thus, the inclusion of this band is made for the sake of
completeness, otherwise it could be ignored. However, the unperturbed proton $S$-band provides valuable information regarding the dynamics of the system's angular momenta. Indeed, the minimum displayed by both the neutron and proton $S$-bands in Fig.\ref{etot} indicates the amount of angular momentum carried by the corresponding broken pair. This is suggested by the following reasoning. First of all one must note that the slopes of the curves from Fig.\ref{etot} determine the rotational frequencies of the bands. The negative slopes of the neutron and proton $S$-bands at low spins imply a negative rotational frequency which is due to the core that must compensate the already high angular momentum realized by the decoupled broken pair. In the minimum point, where the slope vanishes, the core is no longer rotating and the total angular momentum is coming from the broken pair alone. Thus, the spin at which the $S$-bands show a minimum represents the angular momentum carried by the broken pair. 

Inspecting  Fig.\ref{etot} one finds that for all considered nuclei the neutron broken pair carries almost 8-10 units of angular momentum, while the angular momentum of the proton broken pair is about 6-8$\hbar$. But as we already remarked, the second backbending is due to the crossing of the neutron $S$-band with the neutron-proton $S$-band and not with the proton one. Of course the $4qp$ band associated to two broken pairs, one of neutron and another of proton type has a different structure from a $2qp$ $S$-band. As can be seen from Fig.\ref{etot}, such a band has an extended plateau which means that the total angular momentum is due to the both broken pairs without any core contribution. As a matter of fact the total spin where the plateau ends and the core starts to rotate is equal to the sum of the angular momenta provided by the broken pairs, which is around $J=16$. 
\begin{figure}[h!]
\begin{center}
\includegraphics[width=1\textwidth]{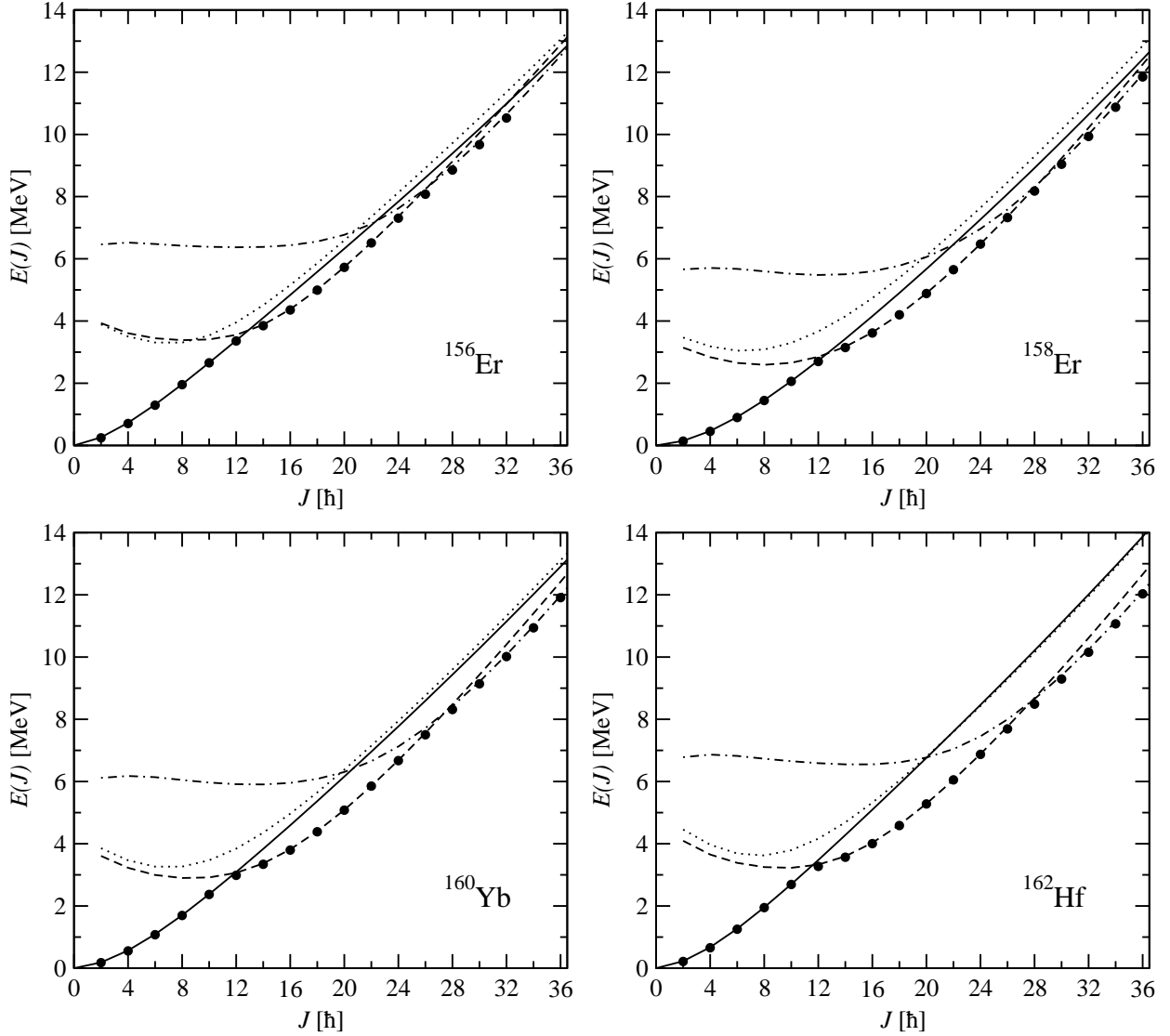}
\end{center}
\caption{Energy trajectories implied in the band mixing are presented as function of the total angular momentum. The $g$-band is represented by the straight line, neutron $S$-band by the dashed line and the proton $S$-band by the dotted line, while the dash-dotted line corresponds to the neutron-proton $S$-band. The yrast energies (circles) resulted from the diagonalization of the total Hamiltonian in the orthogonal basis (\ref{basis}) are also visualized.}
\label{etot}
\end{figure}

\begin{figure}[th!]
\begin{center}
\includegraphics[width=1\textwidth]{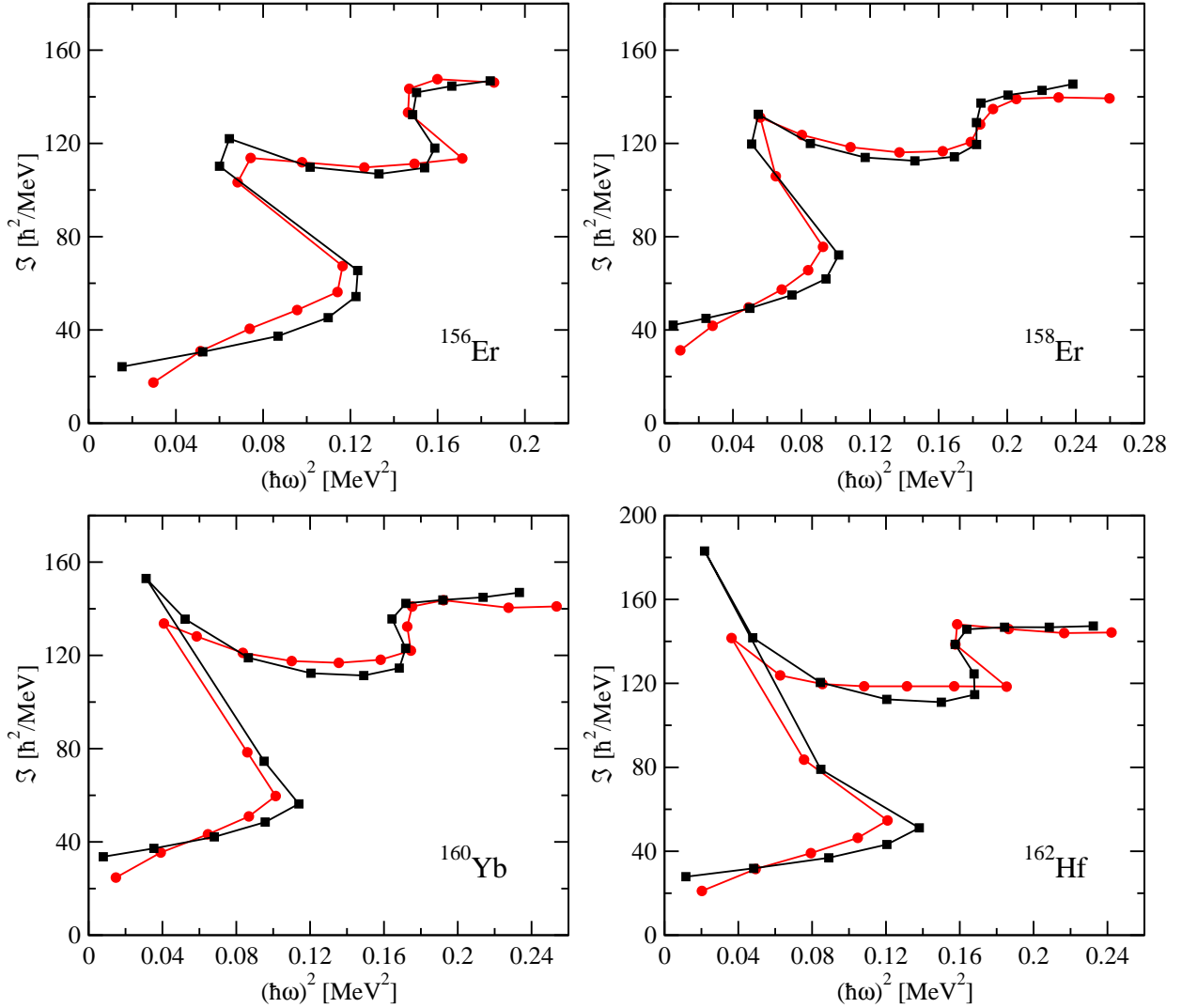}
\end{center}
\caption{Backbending plots for $^{156}$Er, $^{158}$Er, $^{160}$Yb and $^{162}$Hf isotopes comparing theory (squares) with experiment (circles). Experimental data are taken from Refs.\cite{Reich1,Helmer,Reich2,Reich3}.}
\label{bb}
\end{figure}
\label{fijgr}
For a better understanding of the multiple backbending phenomena, the theoretical results and the experimental data are compared by means of backbending plots and the corresponding energy spectra. The backbending plot is a graph which shows the dependence of the moment of inertia on the angular frequency squared. If one adopts for the moment of inertia the following expression
\begin{equation}
\mathfrak{I}=\frac{4J+6}{E(J+2)-E(J)},
\end{equation}
where $E(J)$ are the yrast energies, and defines the rotational frequency as
\begin{equation}
\hbar\omega(J)=\frac{dE(J)}{dJ}\approx\frac{1}{2}[E(J+2)-E(J)],
\end{equation}
one readily obtains the experimental and theoretical backbending curves for the four nuclei treated here. These plots are shown in Fig.\ref{bb} where the description is limited to the experimental yrast states up to the spin $J=36$ for $^{158}$Er, $^{160}$Yb and $^{162}$Hf and $J=32$ for $^{156}$Er. The nature of states with angular momentum higher than 36 might be different from that of the states considered in the present work. Indeed, since the states density increases with the spin, one expects that a larger band admixture takes place. Even so, the number of experimental states described here is enough to account for the most important features of the second moment of inertia anomaly. The smaller number of yrast states considered in the case of $^{156}$Er is due to the fact that the states  beyond $J=32$ have not yet an angular momentum assigned.

Coming back to the backbending plots of Fig.\ref{bb}, it is obvious that the double zigzag shape is reproduced quite well for all four nuclei. An especially good agreement is found for moderate spin states at the first backbending which is, indeed, very well reproduced in all cases. The second backbending is supposed to be less pronounced than the first one, because, as Fig.\ref{etot} shows, the crossing angle between the neutron $S$-band and the neutron-proton $S$-band is much smaller than the one between the $g$-band and the neutron $S$-band. However, the experimental data offers a rather sharp second backbending for nuclei $^{156}$Er, $^{160}$Yb and $^{162}$Hf, while the theoretical calculations predict a  smoother  backbending behavior. In the case of $^{158}$Er, the second observed moment of inertia anomaly is not a real backbending but a relatively weak up-bending. Note that, the theoretical results also predict an up-bending which is however much steeper. This is consistent with the results from Fig.\ref{etot} where the crossing angle between the neutron and neutron-proton $S$-bands for this nucleus is very small.

Concerning the comments on the pairing constants given in subsection B, the question which certainly arises is {\it what is the effect on the backbending plot when we use the $G_{\tau}$
parameters obtained  by fitting the even-odd mass difference instead of the interaction strengths
fixed as described above}. The answer is given  in Fig.\ref{bbYb}, for $^{160}$Yb. Indeed, comparing the results corresponding to the two sets of $G_{\tau}$, one may state that there is no significant difference between the two plots.
\begin{figure}[h!]
\vspace*{0.8cm}
\begin{center}
\includegraphics[width=0.6\textwidth]{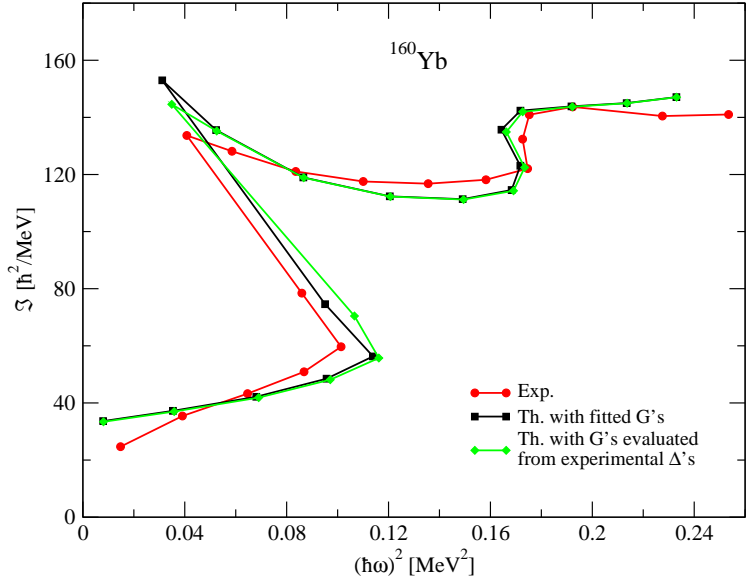}
\end{center}
\caption{The backbending plot corresponding to two sets of parameters for the  pairing strengths: one obtained by the fitting procedure described in the text (squares) and one fixed so that the even-odd mass difference (diamonds) be reproduced.}
\label{bbYb}
\end{figure}
Concluding, the real number of the free parameters is four. 

The good agreement between theoretical and experimental backbending plots is reflected also in the corresponding energy spectra. Thus,  Fig.\ref{spectru} suggests  a very good agreement between the results of our calculations and the corresponding data, which is quantitatively expressed by relatively small root-mean-square ($r.m.s.$) values for deviations. Note that the energy spectra are better reproduced at high spins than  at low spins, contrary to the backbending plots where the first backbending is better described than the second one. This happens because the backbending curves do not depend on the absolute energies of the angular momentum states, but on the energy difference between consecutive states and moreover through a quadratic law $(\hbar\omega)^{2}$ which is more sensitive to small deviations. Examining Fig.\ref{spectru}, one remarks an increasing behavior of the critical energies with $Z$ for the $N=90$ isotones. This feature might be ascribed to the constant decrease of the deformation which increases the frequency of the collective rotation.
\begin{figure}[h!]
\begin{center}
\includegraphics[width=1\textwidth]{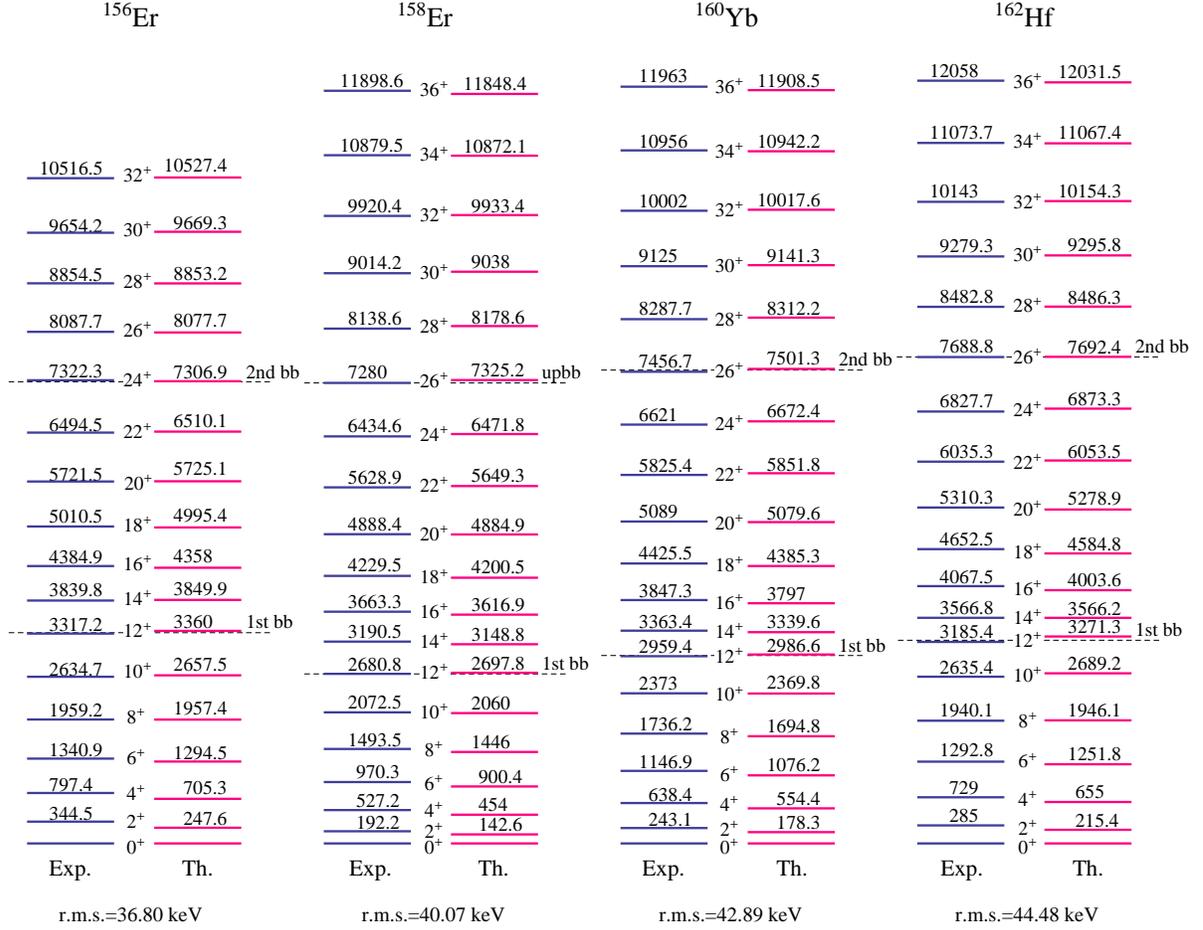}
\end{center}
\caption{Experimental and theoretical yrast spectra of $^{158}$Er, $^{160}$Yb, $^{162}$Hf and $^{156}$Er, with numerical values given in units of keV. The starting point of the backbendings are indicated for each nuclei by a dashed line. At the beginning of each spectrum one can find the corresponding root-mean-square ($r.m.s.$) values.}
\label{spectru}
\end{figure}

The four nuclei treated here are $\gamma$-unstable. Thereby the collective motion of the $N=90$ isotones $^{160}$Yb and $^{162}$Hf can be well described by the $O(6)$ dynamic symmetry \cite{Iachello1}. The softness of these nuclei points to a possible dynamic deformation which is increasing with the angular momentum. Indeed, judging by the behavior of the $g$-bands from Fig.\ref{etot}, the energy spectrum at lower spins is of the rotational type, while for larger spins it becomes more vibrational-like. This change in the energy spectra is most likely caused by the increase of the $\gamma$ deformation because the $\beta$ is fixed for these nuclei. The structure of $^{156}$Er is different. The observed collective spectrum of the $^{156}$Er exhibits signatures of the $E(5)$ dynamical symmetry \cite{Iachello2} which is assigned to the critical point of the phase transition from the $O(6)$ to the $U(5)$  symmetry. The critical point potential has a very extended minimum in the deformation parameter $\beta$ around the origin which corresponds to a spherical shape described by the $U(5)$ dynamical symmetry. As a matter of fact, the observed nuclear deformation of $^{156}$Er is indeed small. In this case one can also have a variation of the $\beta$ deformation along its flat minimum as the nucleus is increasing its rotation. 
\subsection{Angular momentum alignment}
In order to study the alignment of the angular momenta involved in the system's dynamics,  it is useful to compute the averages of the involved angular momenta:
\begin{eqnarray}
\tilde{J}_{n}(\tilde{J}_{n}+1)&=&\langle\Phi_{Tot}^{JM}|\vec{J}_{n}^{2}|\Phi_{Tot}^{JM}\rangle,\label{Jn}\\
\tilde{J}_{p}(\tilde{J}_{p}+1)&=&\langle\Phi_{Tot}^{JM}|\vec{J}_{p}^{2}|\Phi_{Tot}^{JM}\rangle,\\
\tilde{J}_{f}(\tilde{J}_{f}+1)&=&\langle\Phi_{Tot}^{JM}|\vec{J}_{f}^{2}|\Phi_{Tot}^{JM}\rangle,\label{Jf}\\
\tilde{J}_{c}(\tilde{J}_{c}+1)&=&\langle\Phi_{Tot}^{JM}|\vec{J}_{c}^{2}|\Phi_{Tot}^{JM}\rangle.\label{Jc}
\end{eqnarray}
The deviation
\begin{equation}
\Delta J=\left|J-(\tilde{J}_{c}+\tilde{J}_{f})\right|,
\end{equation}
is a measure for the departure from the full alignment of the fermionic and core angular momenta, i.e. when $\tilde{J}_{c}+\tilde{J}_{f}$ equates the total angular momentum $J$ of the system. All the average angular momenta (\ref{Jn})-(\ref{Jc}) and the deviation $\Delta J$ are plotted in Fig.\ref{aligntotal} as the functions of total angular momentum $J$. These plots reveal additional information for  the backbending phenomenon. Indeed, from Fig.\ref{aligntotal} one can extract the angular momentum carried by the neutron and proton broken pairs, the composition of the total angular momentum, the critical spins of the band crossings, or one can even investigate the alignment of different angular momenta of the system. The difference between the values of the $\tilde{J}_{f}$, $\tilde{J}_{n}$ and $\tilde{J}_{p}$ before and after the critical angular momenta associated to the pair breaking, gives the amount of angular momentum carried by the broken pairs which is consistent to those determined from analyzing the plots of Fig.\ref{etot}. An interesting feature can be seen from Fig.\ref{aligntotal}, which is the essential difference between the two band crossings. Indeed, while the neutron angular momentum $\tilde{J}_{n}$ has a clear discontinuity reflected in a jump to a plateau of higher spin, the proton angular momentum has a steady increase extended around the critical angular momentum where the second band crossing actually takes place, although the curve changes substantially its slope. This was somehow expected due to the smaller crossing angle between the neutron and neutron-proton $S$-bands. The smaller crossing angle means a larger range of the angular momentum where the bands are effectively interacting. The neutron and neutron-proton $S$-bands start to interact from $J=22$ for $^{156}$Er and $J=24$ for the rest of nuclei, and keep interacting afterwards. After this spin, the states are no longer of a pure nature and the nucleus is described by a coexistence of $2qp$ and $4qp$ states of broken pairs. This is contrary to the case of the first band crossing where the interacting range is finite and very short, about 2 units of angular momentum. Investigating the behavior of the core angular momentum $\tilde{J_{c}}$, one observes that it has a sudden fall at the first band crossing of about $2\hbar-3\hbar$, while at the second band crossing it drops very little (under $1\hbar$), keeping approximately the same value for few total angular momentum states. 

\begin{figure}[h!]
\begin{center}
\includegraphics[width=1\textwidth]{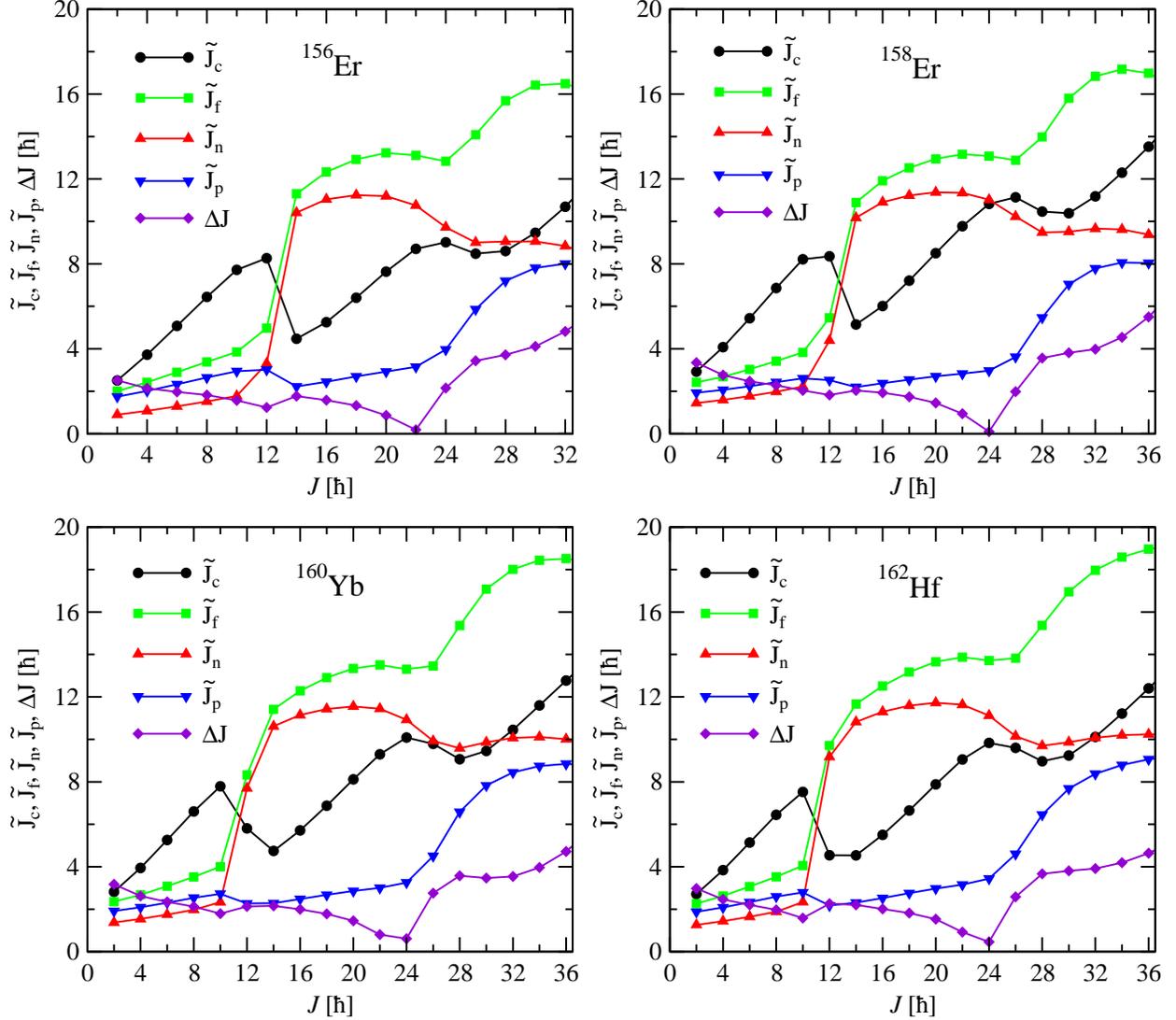}
\end{center}
\caption{Color online. Expected values of the angular momenta corresponding to the neutron and proton broken pairs, the total fermionic angular momentum of the neutron and proton intruder orbitals as well as the core angular momentum. The deviation $\Delta J$ of the total angular momentum is also visualized.}
\label{aligntotal}
\end{figure}

Concerning the angular momenta alignment, one remarks that before the band crossings the alignment defect $\Delta J$ has a minimum and right after a local maximum. Note that here we deal with a rotational alignment and that is why the deviation $\Delta J$ decreases with total angular momentum. Even though, the full alignment $\Delta J=0$ is not possible because of the fact that after the first band crossing the yrast states are of $K\neq0$ nature. However, at the beginning of the second band crossing, one finds that $\Delta J\approx0$. This approximate alignment is due to the fact that the proton orbital starts to aid more consistently the fermionic angular momentum $\tilde{J}_{f}$ when the neutron $S$-band starts to interact with the neutron-proton $S$-band and the proton pair just slowly begins to break. This leads us to the conclusion that the angular momenta of the broken pairs first align to each other and only after that they align with the core angular momentum. The last alignment seems to be hindered, as shown in Fig.\ref{aligntotal} where the angular momentum defect does not decrease after the second band crossing and moreover at some point it starts to increase in parallel with the core angular momentum. The increasing behavior of $\Delta J$ at high angular momentum states points to the fact that the rotation at high spins starts to work against the alignment between the core and fermionic angular momenta.

\subsection{Electric quadrupole transitions}

A very sensitive test of the wave functions describing the energy levels are the quadrupole transition probabilities. In Fig.\ref{tranz} one compares the numerical results provided by the formulas from Sec. III with the corresponding experimental data available only for $^{156}$Er, $^{158}$Er and $^{160}$Yb. The parameters $q_{1}$ and $q_{2}$ of the quadrupole transition operator are fixed by fitting the experimental $B(E2)$ values and the obtained results are given in Table III. The theoretical and experimental values are also compared with the rotational limit of the quadrupole transition probability corresponding to the rigid rotor wave functions defined as:
\begin{equation}
B(E2,J^{+}\,\rightarrow\,J'^{+})_{rot}=\frac{5}{16\pi}Q_{0}^{2}\left(C^{J2J'}_{0\,0\,0}\right)^{2},
\end{equation}
with $Q_{0}$ fixed by fitting the first experimental transition probability $B(E2,0^{+}\,\rightarrow\,2^{+})$. The values of $Q_{0}$ corresponding to each considered nucleus are also given in the Table III.

\begin{table}[h!]
\caption{The results of the fitting procedure performed for the quadrupole transition probabilities shown in Fig.\ref{tranz} are listed for each treated nucleus together with the $Q_{0}$ value defining the values of $B(E2)_{rot}$.}
\begin{tabular}{|c|c|c|c|c|}
\hline
Nucleus&$q_{1}$ [W.u.]$^{1/2}$&$q_{2}$ [W.u.]$^{1/2}$&$r.m.s.$ [W.u.]&$Q_{0}$ [W.u.]$^{1/2}$\\
\hline
$^{156}$Er&12.31060&11.73900&60.3801&57.4669\\
$^{158}$Er&~2.10613&-1.58305&74.4017&77.34075\\
$^{160}$Yb&~7.89159&~7.55747&55.0754&68.37170\\
\hline
\end{tabular}
\end{table}

\begin{figure}[h!]
\begin{center}
\includegraphics[width=0.65\textwidth]{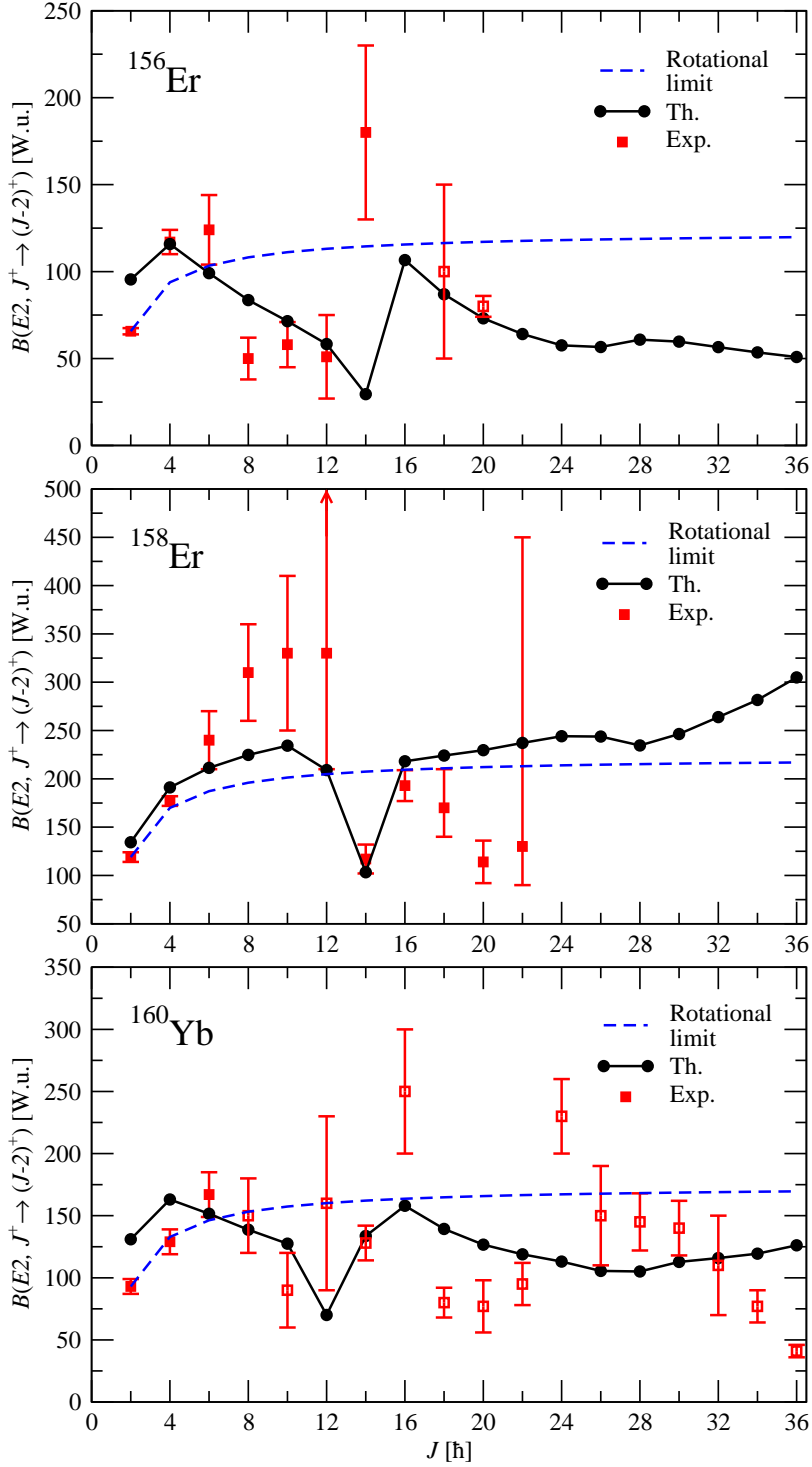}
\end{center}
\caption{Theoretical predictions for the reduced $E2$ transition probabilities are compared with experimentally available data for $^{156}$Er, $^{158}$Er and $^{160}$Yb taken from Refs.\cite{Reich1,Helmer,Reich2}. The open symbols indicate experimental data with assumed or derived assignment and were not taken into account for the fitting procedure only in case of $^{156}$Er nucleus. The rigid rotor limit of the $B(E2)$ is also shown for comparison.}
\label{tranz}
\end{figure}

The transitions along the yrast band directly reflect the structural changes of the total wave function in the band crossing region. Indeed, investigating the theoretical points from Fig.\ref{tranz} one notices that at the first band crossing only one transition is sizably hindered. This indicates the fact that the interaction of the $g$-band with the neutron $S$-band is weak such that the transition from $0qp$ to the $2qp$ nature is very sudden, taking place in the interval of no more than 2 units of total angular momentum. This behavior is also found in the experimental data, although in case of the $^{156}$Er nucleus the minimum calculated transition is somehow shifted to the next transition in respect to experimental results. The situation at the second band crossing is essentially different because in this case both model states are of the quasiparticle nature which enforces the interband interaction leading to a less visible decrease of the transition probability with an extended minimum in the band crossing region. Looking at the experimental values, especially those before the first band crossing, we observe some significant deviations from the rigid rotor behavior. The largest deviations are obtained in case of the $^{156}$Er nucleus. Judging by the moderately small values of the nuclear deformation $\beta_{2}$ and of the obtained values for the deformation parameter $d$, it is not surprising that $^{156}$Er deviates the most from the perfect rigid rotor case. The large discrepancy at the low spins between the experimental data and the predicted rigid rotor behavior could also be due to the fact that these nuclei are relatively sensitive to the shape fluctuations. This is, in fact, consistent with the previous comment about the $\gamma$ softness of these nuclei.  The oscillation of the transition probabilities before the first band crossing, although not yet well understood from the phenomenological point of view, it is well reproduced by the theoretical results. Indeed, even the unusual parabolic dependency on the angular momentum of the $B(E2)$ values before the first band crossing in the $^{158}$Er and $^{160}$Yb nuclei is simulated quite well by the model predictions.

Another feature which deserves attention consists of that the parameters $q_1$ and $q_2$ for 
$^{156}$Er and $^{158}$Er are quite different. One reason was already mentioned, namely that the two isotopes have different deformation which makes $^{158}$Er be closer to the rotor behavior. Another reason might be the fact that $^{156}$Er reaches the conditions of a critical point of shape phase transition $U(5)\to O(6)$ exhibiting a $E(5)$ symmetry which results in having a discontinuity in the strength parameters of the transition operator. Indeed, the ratios
of the excitation energies $E_{4^+}/E_{2^+}$ for the two isotopes  are 2.315 and 2.743, respectively which have to be compared with the $E(5)$ limit which amounts of 2.2. Therefore a smooth behavior of the $q_1$ and $q_2$ parameters in the isotopic chain of Er isotopes is expected to be broken at $^{158}$Er which is close to the critical point of the shape phase transition.

\subsection{Gyromagnetic factor}

The magnetic dipole moment of the particle-core system is defined as:
\begin{equation}
\vec{\mu}=g_{c}\vec{J}_{c}+g_{f}\vec{J}_{f}\equiv g_{J}\vec{J},
\end{equation}
where $g_{c}$ and $g_{f}$ denote the gyromagnetic factors of the core and fermionic subsystems, respectively. The structure of the  total wave function is reflected by the total gyromagnetic factor $g_{J}$:
\begin{equation}
g_{J}=g_{c}+\frac{g_{f}-g_{c}}{2}\left[1+\frac{\tilde{J}_{f}(\tilde{J}_{f}+1)-\tilde{J}_{c}(\tilde{J}_{c}+1)}{J(J+1)}\right].\label{gJ}
\end{equation}
For the core gyromagnetic factor one takes the rotational  value
\begin{equation}
g_{c}=\frac{Z_{c}}{A_{c}},
\end{equation}
given in units of nuclear magneton $\mu_{N}$, where $Z_{c}$ and $A_{c}$ are the nuclear charge  and the mass number of the core:
\begin{eqnarray}
Z_{c}&=&Z-2\left\langle N_{pair}^{\pi h_{11/2}}\right\rangle,\\
A_{c}&=&A-2\left\langle N_{pair}^{\nu i_{13/2}}\right\rangle-2\left\langle N_{pair}^{\pi h_{11/2}}\right\rangle,
\end{eqnarray}
with the expected number of neutron and proton pairs determined from the BCS equations and given in Table II.

\begin{figure}[th!]
\begin{center}
\includegraphics[width=1\textwidth]{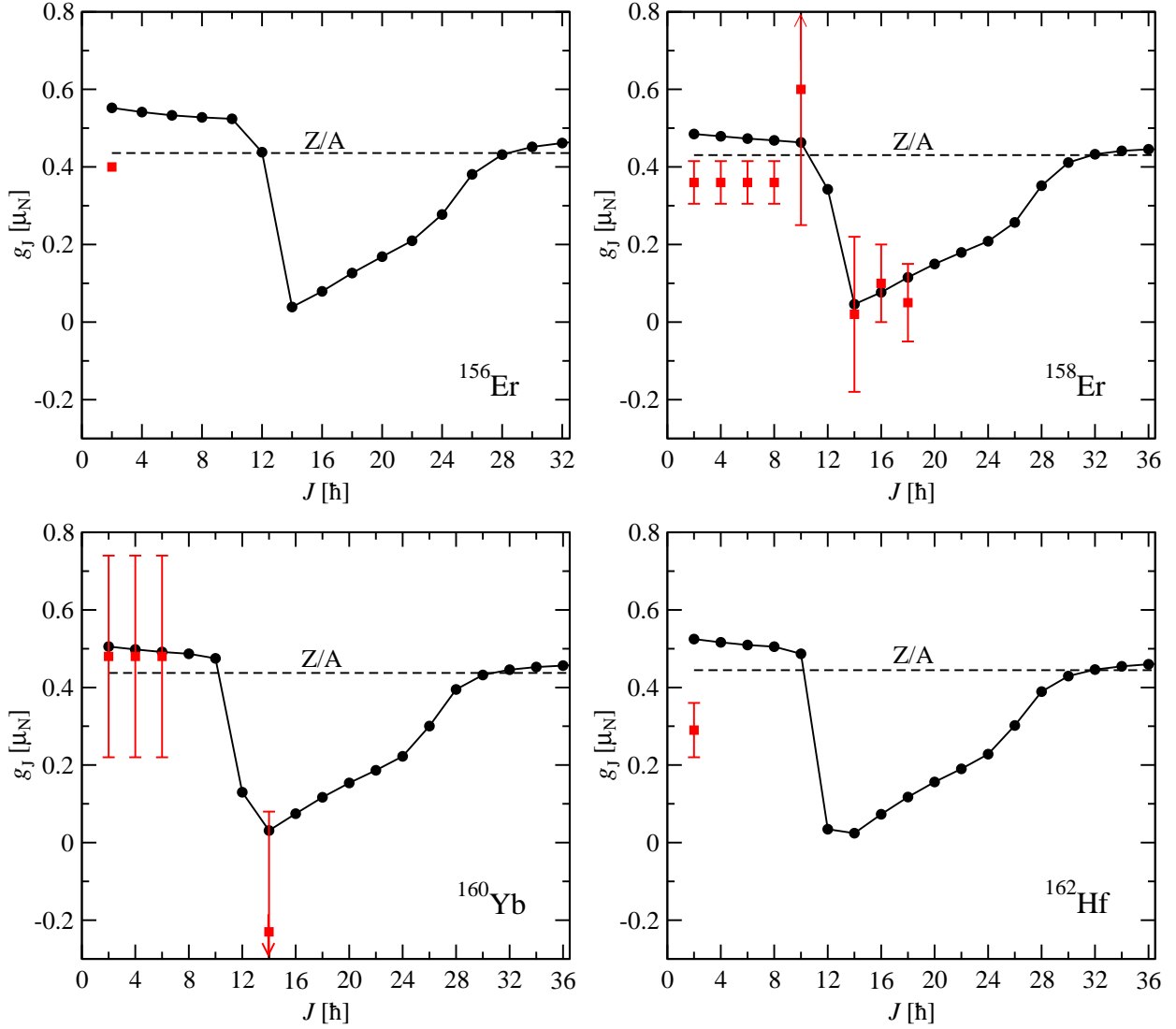}
\end{center}
\caption{Color online. Calculated gyromagnetic factor for yrast states (circles) given in units of nuclear magneton is represented as function of angular momentum. There are also visualized few experimental values (squares) taken from Refs.\cite{Reich1,Helmer,Reich2,Reich3} together with the rotational limit $Z/A$ of the gyromagnetic factor.}
\label{gyro}
\end{figure}

As for the fermionic gyromagnetic factor, it is obtained from the following decomposition of the fermionic  magnetic moment:
\begin{equation}
\vec{\mu}_{f}=g_{f}\vec{J}_{f}=g_{n}\vec{J}_{n}+g_{p}\vec{J}_{p},
\end{equation}
which gives an expression for $g_{f}$ in terms of $\tilde{J}_{n}$, $\tilde{J}_{p}$ and $\tilde{J}_{f}$ similar to (\ref{gJ}),
\begin{equation}
g_{f}=g_{p}+\frac{g_{n}-g_{p}}{2}\left[1+\frac{\tilde{J}_{n}(\tilde{J}_{n}+1)-\tilde{J}_{p}(\tilde{J}_{p}+1)}{\tilde{J}_{f}(\tilde{J}_{f}+1)}\right].
\end{equation}
Knowing that the intruder neutrons are from the $i_{13/2}$ orbital, and the intruder protons are from the $h_{11/2}$ orbital, one obtains the following values for the proton and neutron gyromagnetic factors
\begin{eqnarray}
g_{n}&=&g^{(n)}_l+(g^{(n)}_s-g^{(n)}_l)/13=\frac{g_{s}}{13}=-0.22\mu_{N},\\
g_{p}&=&g^{(p)}_l+(g^{(p)}_s-g^{(p)}_l)/11=1.29\mu_{N}.
\end{eqnarray}
For the above calculation we used the free value of the $g_{l}$ while for $g_{s}$ the free values  were quenched by the factor 0.75, which accounts for the nuclear medium effect \cite{Castel},
\begin{equation}
g_{l}^{n}=0,\,\,\,g_{l}^{p}=1\,\mu_{N},\,\,\,g_{s}^{n}=-3.8256\times0.75\,\mu_{N},\,\,\,g_{s}^{p}=5.5855\times0.75\,\mu_{N}.
\end{equation}

The total gyromagnetic factor is plotted in Fig.\ref{gyro} as function of the total angular momentum $J$. Its change in the behavior reflects the transition from states of different nature. Before the first band crossing its value is almost constant and close to the rotational limit, although slightly overestimated. Of course, even if the nature of the $g$-band is collective, it is far from being perfectly rotational as it is suggested by the small values of the deformations $d$ and $\beta_{2}$ from Table I. Indeed, it can be seen from Fig.\ref{gyro} that the departure of the gyromagnetic factor from its rotational limit $Z/A$ before the first band crossing is bigger for $^{156}$Er and $^{162}$Hf nuclei, which turn out to be the less deformed ones. At the first band crossing the gyromagnetic factor has a sudden fall down, reaching very small values where the total magnetic moment almost vanishes. This discontinuity marks the change of the yrast band from $0qp$ to a $2qp$ neutron character. The fall of $g_{J}$ at the first band crossing is due to the negative value of the neutron gyromagnetic factor coming from the decoupled neutron pair. After the first band crossing the rotation of the core starts to dominate and the gyromagnetic factor increases almost linearly with $J$. This trend keeps up to the second band crossing where the ascendant slope becomes bigger due to the positive value of the proton gyromagnetic factor coming from the proton broken pair. The second band crossing is reflected in an inflexion point of $g_J$ as function of $J$. This is consistent to the slowness of the consequent breaking of the proton pair which does not offer a jump like in the case of the first band crossing. The growth of the $g_{J}$ persists only for a few states and then it comes to a saturation in the vicinity of the rotational limit value. As a matter of fact, the mentioned plateau begins at the spin where the second backbending ends. Few remarks are necessary regarding the comparison of calculation results with the experimental values of the gyromagnetic factor. Leaving aside the nuclei $^{156}$Er and $^{162}$Hf where relevant experimental data are lacking, the other two reproduce quite well the sudden fall of $g_{J}$ at the first band crossing. An especially good agreement between theory and experiment is obtained for $^{158}$Er where not only the discontinuity of the gyromagnetic factor but also its absolute values are reproduced.

Before closing this section, we would like to comment on the obtained values of some of the model parameters. First of all, one notes the linear dependence of the deformation parameter $d$ on the nuclear deformation $\beta_{2}$. This property can be used to approximately determine the deformation $d$ for other nuclei from the rare earth region. The numerical values of the deformation parameter $d$ are in the range of values determined in Refs.\cite{Rabura1,Rabura2} for other isotopes of the nuclei treated in this paper. This feature pleads in favor of both the CSM formalism and the present approach. The other parameter which deserves a special attention is the strength of the spin-spin interaction. Although such an interaction was already used in connection to the backbending phenomena \cite{Ikeda1}, here it brings  an essentially different contribution. First of all in Ref.\cite{Ikeda1}, the spin-spin interaction was found to be repulsive while in the present model it can be both attractive and repulsive. Indeed, the spin-spin interaction matrix elements are going from negative to positive values in the $2qp$ and $4qp$ bands as well as in the corresponding non-diagonal matrix elements. The picture is opposite for negative values of the strength $C$, as  happens in the case of $^{162}$Hf. It is interesting to mention that the second backbending in $^{162}$Hf, is difficult to explain due to its unexpected sharpness. Indeed, before the second backbending was experimentally observed in $^{162}$Hf, the CHFB  calculations predicted for this nucleus a small up-bending or even no backbending \cite{Faessler}. As a matter of fact in our approach the reproduction of the second backbending in this nucleus was possible only by choosing a negative value for the spin-spin strength $C$. This feature proves the importance of the spin-spin interaction in explaining the backbending phenomenon which thus appears to be the result of an interplay between the Coriolis-like force and the $Qq$ interaction.

\renewcommand{\theequation}{7.\arabic{equation}}
\section{Conclusions}
\label{sec: level7} 

The present model  provides a consistent explanation for the pair breaking process in connection to the rotational alignment of the angular momenta involved in the system.  Using simple arguments one determines the critical angular momentum $J$ where the pair breaking takes place.  While the neutron pair breaking takes place at $J=10$ or $12$, one cannot accurately say at what angular momentum the proton pair is broken because at high spin states the crossing bands interact within a larger interval. This is suggesting that the proton pair breaking is a slower process than the neutron pair breaking.

Concerning the rotational alignment, it is found that the proton and neutron angular momenta first align to each other and only after that they align to the core angular momentum. The full alignment between the fermionic and the core angular momenta cannot be achieved due to the intrinsic properties of the higher spin states which are of the $K=1$ and $K=2$ nature. However, strong alignments are obtained at the band crossing critical angular momenta. Another interesting result of the present approach is that the rotational alignment lessens after the second backbending, which is pointing to the fact that the $2qp$ and $4qp$ bands still interact even after the band crossing.

The first backbending manifests itself in the gyromagnetic factor plot by a big fall down of $g_J$.
By contradistinction, the second backbending is reflected by an inflexion point in the above mentioned plot.

The effect brought by each term of the model Hamiltonian on the spectrum in the region of the band crossing is in extenso analyzed. In this way the free  parameters acquire a well established significance. 

Along the time, various versions of angular momentum projection has been used with the aim of describing the backbending phenomena
\cite{Faessler2,Beng2,Grum,Lee,Faessler1,Plosz,Li,Egido1,Egido2,Ikeda1}.

What distinguishes our model from the others? First of all the three components of neutrons, protons and the core are described by deformed wave functions. Moreover, the mean fields of neutrons as well of protons are derived from the particle-core coupling term and thereby the three components have similar deformation properties. The total wave function
describing the nucleus in the laboratory reference frame is obtained by angular momentum projection procedure from the product of the mentioned three deformed functions which,  is not an easy task. We suspect that due to the specific construction, the wave function has a complex structure which allows to describe quantitatively the spectra in the region of the two backbendings. The accuracy of description is reflected not only in the backbending plot but also by transition probabilities (Fig.9) and gyromagnetic factors (Fig.10).

Note that the core is described by projecting out the angular momentum from a coherent state and by an anharmonic boson Hamiltonian. Therefore the  core moment of inertia is not constant but depending on the angular momentum. In this context we could assert that our model is on a par with those particle-core approaches using a variable moment of inertia.

As a final conclusion one can say that the present formalism is able to describe quantitatively the double backbending phenomenon. Moreover,  a consistent qualitative explanation of the combined contribution of the pair breaking and rotational alignment to the backbending phenomenon is provided.

{\bf Acknowledgment.} This work was supported by the Romanian Ministry for Education Research Youth and Sport through the CNCSIS project ID-2/5.10.2011.

\end{document}